\documentclass{aa}  

\usepackage{graphicx}
\usepackage{txfonts}
\usepackage[colorlinks=true]{hyperref}
\hypersetup{linkcolor=blue, citecolor=blue}
\usepackage[draft]{todonotes}
\usepackage{mathtools}
\usepackage{booktabs}
\usepackage{multirow}
\usepackage{bbold}
\usepackage{caption}
\usepackage{subcaption}
\usepackage{bm}
\usepackage{ulem}
\usepackage{tikz}
\usepackage{xcolor}
\definecolor{vaso}{rgb}{1.0, 0.0, 0.0}
\definecolor{alex}{rgb}{0.0, 0.0, 1.0}

\newcommand{\B}{\mathbf{B}}
\newcommand{\bfphi}{\bm{\varphi}}
\newcommand{\kbf}{\mathbf{k}}
\newcommand{\xbf}{\mathbf{x}}

\newcommand{\paperI}{Paper I}

\tikzset{viewport/.style 2 args={
    x={({cos(-#1)*1cm},{sin(-#1)*sin(#2)*1cm})},
    y={({-sin(-#1)*1cm},{cos(-#1)*sin(#2)*1cm})},
    z={(0,{cos(#2)*1cm})}
}}

\newcommand{\ToXYZ}[2]{
    {sin(#1)*cos(#2)}, % X coordinate
    {cos(#1)*cos(#2)}, % Y coordinate
    {sin(#2)}          % Z (vertical) coordinate
}

\def\RotationX{-30}
\def\RotationY{-20}

\begin{document}

   \title{Non-parametric Bayesian reconstruction of Galactic magnetic fields using Information Field Theory}

   \subtitle{The inclusion of line-of-sight information in ultra-high energy cosmic ray backtracking}

    \titlerunning{Non-parametric Bayesian reconstruction of GMFs}

   \author{Alexandros Tsouros \inst{1} \fnmsep \inst{2}\thanks{tsouros@physics.uoc.gr}, Abhijit B. Bendre \inst{3}\fnmsep\inst{4}, Gordian Edenhofer  \inst{5}\fnmsep\inst{6}\fnmsep\inst{7}, Torsten Enßlin \inst{5}\fnmsep\inst{6}, Philipp Frank \inst{5}, Michalis Mastorakis \inst{1} \fnmsep \inst{2} , Vasiliki Pavlidou \inst{1} \fnmsep \inst{2}
          }

   \institute{
        Department of Physics \& ITCP, University of Crete, GR-70013, Heraklion, Greece
        \and
        Institute of Astrophysics, Foundation for Research and Technology-Hellas, Vasilika Vouton, GR-70013 Heraklion, Greece
        \and Laboratoire d’Astrophysique, EPFL, CH-1290 Sauverny, Switzerland
        \and Scuola Normale Superiore di Pisa, Piazza dei Cavalieri 7, 56126 Pisa, Italy
        \and Max Planck Institute for Astrophysics, Karl-Schwarzschild-Stra{\ss}e 1, 85748 Garching, Germany
        \and Ludwig Maximilian University of Munich, Geschwister-Scholl-Platz 1, 80539 Munich, Germany
        \and University of Vienna, Department of Astrophysics, Türkenschanzstrasse 17, 1180 Vienna, Austria
        }
    \authorrunning{A. Tsouros et al. 2023}
    
   \date{Received ; accepted }

\abstract
  % context heading (optional)
  % {} leave it empty if necessary  
   {Ultra-high energy cosmic rays (UHECRs) are extremely energetic charged particles with energies surpassing $10^{18}$ eV. Their sources remain elusive, obscured by deflections caused by the Galactic magnetic field (GMF). This challenge is further complicated by our limited understanding of the three-dimensional structure of the GMF, as current GMF observations consist primarily of quantities integrated along the line-of-sight (LOS). Nevertheless, data from upcoming stellar polarisation surveys along with Gaia's stellar parallax data are expected to yield local GMF measurements.}
  % aims heading (mandatory)
   {This study is the second entry in our exploration of a Bayesian inference approach to the local GMF that uses these forthcoming local GMF measurements, by attempting to reconstruct its $3$D structure. The ultimate aim is to backtrack observed UHECRs, thereby updating our knowledge about their possible origin.}
  % methods heading (mandatory)
   {We employ methods of Bayesian statistical inference in order to sample the posterior distribution of the GMF within part of the Galaxy. By assuming a known energy, charge, and arrival direction of an UHECR, we backtrack its trajectory through various GMF configurations drawn from the posterior distribution. Our objective is to rigorously evaluate our algorithm's performance in scenarios that closely mirror the setting of expected future applications. In pursuit of this, we condition the posterior to synthetic integrated LOS measurements of the GMF, in addition to synthetic local POS-component measurements. In this proof of concept work, we assume the ground truth to be a magnetic field produced by a dynamo simulation of the Galactic ISM.}
  % results heading (mandatory)
   {Our results demonstrate that for all locations of the observed arrival direction on the POS, our algorithm is able to substantially update our knowledge on the original arrival direction of UHECRs with rigidity $E/Z = 5 \times 10^{19}$ eV, even in the case of complete absence of LOS information. If integrated data is included in the inference, then the regions of the celestial sphere where the maximum error occurs diminishes greatly. Even in those regions the maximum error is diminished by a factor of about $3$ in the specific setting studied. Additionally, we are able to identify the regions where the largest error is expected to occur.}
  % conclusions heading (optional), leave it empty if necessary 
   {}

   \keywords{Galactic magnetic field --
               Ultra high energy cosmic ray sources --
                Interstellar turbulence
               }

   \maketitle
%
%-------------------------------------------------------------------
\section{Introduction}

Determining the origins of ultra-high-energy cosmic rays (UHECRs) is a crucial challenge in the field of high-energy astrophysics. Successfully addressing this challenge could offer insights with regard to astrophysical processes responsible for generating UHECRs, as well as their composition. Additionally, knowledge of UHECR sources would be a crucial ingredient in multi-messenger studies of high-energy systems (e.g. \citealt{Murase}; \citealt{multimessenger}).

Although numerous theoretical models have been proposed to explain the sources of UHECRs (e.g \citealt{bhattacharjee}; \citealt{TorresReview}; \citealt{KoteraOlinto}), pinpointing these sources has proven to be a complicated task. The main challenge arises from the fact that UHECRs are charged particles, and are deflected by both the Galactic magnetic field (GMF) and the intergalactic magnetic field. As a result, even if multiple UHECRs were emitted from a single, intense, and proximate cosmic ray source (\citealt{Auger-TA}), their trajectories would be dispersed across the plane of the sky (POS). Consequently, any UHECR hotspot would not align with the source. Rather, it would be displaced away from it due to systematic deflections by the ordered component of the GMF, in addition to being spread out due to the random deflections due to the turbulent component of the GMF. This situation contrasts with that of photons or neutrinos, where establishing a connection between observed events and their probable sources is more straightforward, even in the limit of low statistics and poor angular resolution of their detectors.

The primary challenge in understanding the GMF lies in the difficulty of obtaining three-dimensional tomographic reconstruction of the intervening GMF, as the majority of the currently accessible observations are integrated along the LOS. This limitation has guided the predominant approach in GMF modelling to rely on parametric models. This is typically achieved by fitting parameters to distinct analytic components, e.g. a toroidal component, a poloidal component, and a turbulent component. For modelling the latter, a Gaussian random field is employed (\citealt{Sun10}; \citealt{Sun10-2}; \citealt{Takami}; \citealt{JF12}; \citealt{JF12-2}).

However, direct insights into the three-dimensional structure of the interstellar medium of the Milky way are attainable. The Gaia mission, by accurately measuring stellar parallaxes, has mapped the positions of over a billion stars in the Galaxy (\citealt{Gaia1}; \citealt{Gaia-2}; \citealt{Bailer-Jones}). This dataset, combined with other spectroscopic data, has enabled the construction of three-dimensional tomographic maps showing the dust density distribution in certain regions of the Galaxy (\citealt{Lallement2018}; \citealt{Green}; \citealt{Lallement}; \citealt{LeikeEnsslin};  \citealt{Leike2020}; \citealt{Lallement2022}; \citealt{Leike2022}; \citealt{gordian}). Nevertheless, these maps primarily focus on dust density and do not directly constrain the magnetic field.

Yet, observational methods available that probe the three-dimensional structure of the GMF do exist. A notable example is the linear polarization of starlight. Typically, starlight originates from its source as unpolarized light, but can become linearly polarized due to the dichroic absorption by interstellar dust particles, which align themselves with the surrounding magnetic field (\citealt{Andersson}).

Future optopolarimetric surveys, like PASIPHAE and SouthPol, are poised to deliver high-quality stellar polarization measurements for millions of stars (\citealt{Southpol}; \citealt{Pasiphae}; \citealt{WALOP-south1}; \citealt{WALOP-south2}). When combined with the stellar distance data obtained from the Gaia survey, these measurements will enable direct tomographic measurements of the GMF's POS component in regions where dust clouds are present (\citealt{Davis}; \citealt{Chandra}; \citealt{Panopoulou}; \citealt{ST}; \citealt{ST2}; \citealt{Pelgrims}). Additionally, local information can be obtained through the study of HI gas in different velocity bins, which also provide local GMF information (\citealt{Tritsis2018}; \citealt{Tritsis2019}; \citealt{Clark-Hensley}). This information, in conjunction with available LOS data (see, for example, \citealt{Tahani2022a}; \citealt{Tahani2022b}), promises to provide localized and sparse GMF data in the future. This will be instrumental in creating three-dimensional tomographic maps of specific areas of interest. With such maps it becomes feasible to backtrack the paths of UHECRs through these regions, improving source localization on the sky\footnote{However, the contribution of the intergalactic magnetic field is still not accounted for.}. Specifically, there is an intense interest in mapping the GMF in the direction of UHECR `hotspots', as well as in parts of the Galaxy likely to have been traversed by particles comprising these hotspots (\citealt{Abbasi}; \citealt{AugerHotspot}; \citealt{Kawata}).

This study is the second entry in our effort to reconstruct the GMF non-parametrically in $3$D in a Bayesian setting. It directly follows \citealt{tsouros},  hereafter \paperI. Essentially, we address an inverse problem within a Bayesian framework, where the goal is to sample the posterior distribution of GMF configurations in a specific part of the Galaxy, using a combination of local and LOS-integrated information. In this work, local measurements only provide information for the POS component of the magnetic field. This corresponds to the information content of tomographic measurements of interstellar magnetized dust through optopolarimetry of starlight. On the other hand, LOS-integrated measurements provide information for the LOS component of the magnetic field as derived for instance from Faraday rotation measurements (\citealt{Pandhi}; \citealt{faraday}). We will tackle this problem within the context of Information Field Theory, which was developed specifically for Bayesian inference for fields and has been applied successfully in various contexts (\citealt{IFT0}; \citealt{Ensslin2019}; \citealt{IFT-AI}). By reconstructing the posterior distribution of GMF realizations, we aim to accurately recover the true arrival directions of UHECRs given the observed arrival directions, accounting for the influence of the GMF. 

In section \ref{sec:methods}, we briefly describe the methodology, the forward models used, and how the posterior is sampled. In section \ref{sec:results} we present the main results of the algorithm for the considered scenarios, and in section \ref{sec:discussion} we discuss the results further.

\section{Methodology} \label{sec:methods}

In general, we are interested in inferring the configuration of the GMF, $\bm{B}(\mathbf{x})$ with $\xbf \in \mathcal{V}$ over a domain $\mathcal{V} \subset \mathbb{R}^3$, given some observed data set $d$. In the context of Bayesian inference for continuous signals, the task is to determine the posterior probability distribution of $\bm{B}(\mathbf{x})$ conditional to $d$:
\begin{equation}\label{Bayes}
  P(\bm{B}|d) = \frac{1}{Z} P(d|\bm{B}) P(\bm{B}).
\end{equation}
Here, $P(d|\bm{B})$ is the likelihood, representing the probability of observing magnetic field measurements $d$ given a specific configuration $\bm{B}(\mathbf{x})$. The prior, $P(\bm{B})$, encapsulates pre-existing information about $\bm{B}(\mathbf{x})$ while $Z = P(d)$ is the normalisation factor.

In this work, the field that serves as a ground truth (the `true' field) is generated from a dynamo MHD simulation discussed in Appendix \ref{sec:groundTruth}. The original simulation domain extended to $\sim 1$ kpc in the $x - y$ direction, and $\sim 2$ kpc above the Galactic plane. The GMF is rescaled so that its root-mean-square (RMS) value is $5 \mu$G.

\subsection{Likelihood} \label{sec:likelihood}

 Tomography of the magnetized ISM from  stellar polarisation measurements is a highly nontrivial problem and its full discussion is beyond the scope of this work (\citealt{Pelgrims}). However, the reader should be aware that through the combination of Gaia data as well as stellar polarization data for stars of known distance from the Sun, it is possible to acquire information on the Stokes parameters that each intervening dust cloud imposes on the observed starlight. This can then be translated into local information on the orientation of the POS component of the GMF at that cloud, through the connection to grain alignment, as referenced briefly in the previous section and thoroughly in \citealt{Pasiphae}. Information on the POS component of GMF in clouds can also be acquired by the use of  $21$ cm neutral hydrogen (HI) emission measurements (\citealt{Clark-Hensley}). In this work, we assume that the task of determining the locations to which the measurements correspond to has been carried out.

Thus, for the $i$-th datapoint, we assume a forward model of the form
\begin{align}
    &\mathbf{d}_\text{local}^{(i)} = \int R_\text{local}(\xbf, \xbf_i)\bm{B} (\xbf) d^3x + \mathbf{n}_\text{local}^{(i)}, \\ 
    &R_\text{local}(\xbf, \xbf_i) \equiv \delta^{(3)}(\xbf - \xbf_i) P_\text{POS}, 
\end{align}
where $\mathbf{B}(\xbf)$ is the magnetic field, and $\mathbf{n}_\text{local}^{(i)}$ are the observational uncertainties that contaminate our measurements. The vector $\xbf_i$ is the location of the $i$-th cloud where the magnetic field is measured, $P_\text{POS}$ signifies a projection operator on the POS,  which reflects that (mainly) the POS component of the magnetic field is measured via dust polarization, $P_{\text{POS},ij} = \delta_{ij} - \hat{x}_i \hat{x}_i ^\text{T}$ with 
$\hat{x}_i=x_i/||x_i||$  (assuming the observer to be at the origin). The Dirac delta function localizes the measurements at specific known locations $\mathbf{x}_i$. 

% To ensure that the magnetic field is divergence free, we assume it related to a non-divergence-free Gaussian process vector field $\bm{\varphi}$ by a divergence cleaning operator $\mathcal{P}$. This transverse projection operator, defined in Fourier space as   

% \begin{equation} \label{projector}
%   \mathcal{P}_{ij}(\mathbf{k}) = \delta_{ij} - \hat{k}_i \hat{k}^\text{T}_j,
% \end{equation}
% projects out the degrees of freedom of the Gaussian random vector field that violate the divergence-free condition. Said differently, it connects a latent field $\bm{\varphi}(\xbf)$ to the true magnetic field by the harmonic space relation 

% \begin{equation} \label{re_def_bf}
% \hat{B}_i(\kbf) = \frac{3}{2} \mathcal{P}_{ij}(\kbf)\hat{\varphi}_j (\kbf),
% \end{equation}
% where $\kbf$ are Fourier modes. Eq. \ref{re_def_bf} ensures that \(\nabla \cdot \B = 0\), while the factor \(3/2\) accounts for power loss due to reduced degrees of freedom, aligned with the original assumption of statistical isotropy for \(\bfphi\) (\citealt{Hammurabi}).

The option to include the operator $P_\text{POS}$ into the considered scenario is central to this work, as it consists one of the main additions compared to \paperI. A complete projection on the POS is a pessimistic scenario, as LOS information can become available by incorporating Zeeman or Faraday rotation data (\citealt{Tahani2022a}; \citealt{Tahani2022b}). A complete projection on the POS should therefore be seen as an extreme benchmarking scenario. 

We note that this forward model is quite simplistic, in that it assumes that accurate 3D locations are measured. Formally, this is captured by the Dirac delta function and that the locations $\mathbf{x}_i$ are to be assumed known. However, as we will see in section \ref{sec:gp}, the resolution of our reconstruction is of the order of tens of parsecs, corresponding to the uncertainty of cloud localisation (\citealt{Pelgrims}).

The vector $\mathbf{n}_{\text{local}}^{(i)}$ is assumed to be a random variable drawn from a Gaussian distribution with a known covariance $N_\text{local}$. Note that once specific measurement techniques are identified, other more appropriate error distributions will be chosen. Marginalizing over the noise, the likelihood becomes
\begin{equation}
    P(\bm{d}|\mathbf{B}) = \mathcal{G}(\bm{d}_\text{local} - R_\text{local}\bm{B},N_\text{local}).
\end{equation}
The covariance $N_\text{local}$ is chosen to be a multiple of the identity, $(N_\text{local})_{ij} = \sigma^2 \delta_{ij}$, where we choose
\begin{equation} \label{noise_cov_local}
        \sigma = \frac{\mathbf{|B|}_\text{RMS}}{2}, 
    \end{equation}
where $\mathbf{|B|}_\text{RMS} = 5 \mu$G is the RMS value of the magnitude of the ground truth. It should be noted that this does not imply that the noise is correlated with the GMF covariance, it is merely chosen as such in order to ensure an SNR of about 2. 

In addition to local data, in this work we explore the possibility of integrated LOS data, as inferred for instance from Faraday measurements (\citealt{faraday}). In this case, the forward model takes the form 

\begin{align}
    &d_\text{int}^{(i)} =  (\overline{P_\text{LOS} \bm{B}})_{L_i}
 + n_\text{int}^{(i)}, \\
 & (\overline{P_\text{LOS} \bm{B}})_{L_i} \equiv \frac{1}{|L_i|}\int_0^{|L_i|} B_{||} (\bm{x}) d \ell,  
\end{align}
where $P_\text{LOS}$ projects a vector onto the LOS component ($B_{||}$), and $L_i$ the specific LOS under consideration. Further, $|L_i|$ denotes the limit up to which we integrate - in this application $|L_i|$ coincides with the distance between the Earth and the intersection of $L_i$ with the boundary of $\mathcal{V}$. Essentially, the above is equivalent to assuming that the electron density is roughly constant and known up to $|L_i|$ and then falls to zero. While this is not a valid assumption for low Galactic latitudes, we will maintain it in this proof-of-concept work. Finally, the vector $n_\text{int}^{(i)}$ corresponds to a random vector on the POS, with covariance $N_\text{int}$. 

The likelihood in this case is given by 

\begin{equation} \label{eq:likelihood}
    P(\bm{d}|\mathbf{B}) = \mathcal{G}(\bm{d}_\text{local} - R_\text{local}\bm{B},N_\text{local}) \mathcal{G}(d_\text{int} - (\overline{P_\text{LOS} \mathcal{P} \bm{B}})_{L_i},N_\text{int}).
\end{equation}
% The covariance $N_\text{local}$ is chosen to be a multiple of the identity, $(N_\text{local})_{ij} = \sigma^2 \delta_{ij}$, where we choose
% \begin{equation} \label{noise_cov_local}
%         \sigma = \frac{\mathbf{|B|}_\text{RMS}}{2}, 
%     \end{equation}
% where $\mathbf{|B|}_\text{RMS} = 5 \mu$G is the RMS value of the magnitude of the ground truth. It should be noted that this does not imply that the noise is correlated with the GMF covariance, it is merely chosen as such in order to ensure an SNR of about 2. 
    
Similarly, we define the covariance for the noise of the integrated measurements as $(N_\text{int})_{ij} = \sigma_{\text{int}}^2 \delta_{ij}$, where\footnote{While Faraday data is significantly more accurate than this assumption suggests, we will use this pessimistic noise covariance to compensate for the unknown $3$D electron density distribution.} 
\begin{equation} \label{noise_cov_int}
    \sigma_{\text{int}} = \frac{1}{2} \mu \text{G}.
\end{equation}
 Finally, the operator $R_\text{local}$, which sparsely samples the GMF, is defined as follows. After discretising our domain to voxels (see section \ref{sec:bias}), we apply a Bernoulli trial to each voxel to determine whether it is observed or not with probability $p$ and $1-p$ respectively. The probability $p$ is given by the expression

\begin{equation} \label{q}
p =
\begin{cases}
3 \times 10^{-3}, & \text{if } T \geq 10^{4} \text{ K} \\
3 \times 10^{-2}, & \text{if } T < 10^{4} \text{ K}
\end{cases}
\end{equation}
where $T$ is that voxel's corresponding gas temperature, acquired from the same simulation that produced our ground truth. This choice of $p$ reflects the decay of the number of dust clouds as a function of distance from the Galactic plane, which directly correlates with the expected number of measurements with respect to the position above the Galactic plane, as the local measurements of the GMF will ultimately exist where dust clouds are located, after polarized-starlight tomography has been carried out. The specific values chosen are such that the resulting density of points within the domain is roughly $100$ measurements per kpc$^3$ on average. 

\subsection{Prior} \label{sec:prior}

As in \paperI, the only hard constrain that needs to be imposed is that whatever candidate magnetic field configuration $\bm{B}$ we consider, it must satisfy $\nabla \cdot \bm{B} = 0$ in order to be a viable candidate. To ensure that the magnetic field is divergence free, we assume it is related to a non-divergence-free random field $\bm{\varphi}$ by a divergence cleaning operator $\mathcal{P}$. This transverse projection operator, defined in Fourier space as   

\begin{equation} \label{projector}
  \mathcal{P}_{ij}(\mathbf{k}) = \delta_{ij} - \hat{k}_i \hat{k}^\text{T}_j,
\end{equation}
projects out the degrees of freedom of the Gaussian random vector field that violate the divergence-free condition. Said differently, it connects a latent field $\bm{\varphi}(\xbf)$ to the true magnetic field by the harmonic space relation 

\begin{equation} \label{re_def_bf}
\hat{B}_i(\kbf) = \frac{3}{2} \mathcal{P}_{ij}(\kbf)\hat{\varphi}_j (\kbf),
\end{equation}
where $\kbf$ are Fourier modes. Eq. \ref{re_def_bf} ensures that \(\nabla \cdot \B = 0\), while the factor \(3/2\) accounts for power loss due to reduced degrees of freedom, aligned with the original assumption of statistical isotropy for \(\bfphi\) (\citealt{Hammurabi}). Our aim is reconstructing the local GMF $\mathbf{B}$ by inferring the latent field $\bm{\varphi}$ which is related to the latter by Eq. (\ref{re_def_bf}). For $\bm{\varphi}$ we will assume a Gaussian prior of the form:
\begin{equation}\label{Gaussian_vec}
  \mathcal{P}( \bm{\varphi}) = \frac{1}{|2\pi \Phi|^{\frac{1}{2}}}\exp \left[- \frac{1}{2}\int d^3 x d^3 x' \sum_{ij}\varphi_i(\mathbf{x}) \Phi_{ij}^{-1} (\xbf, \xbf')\varphi_j(\mathbf{x}') \right].
\end{equation}
The quantity $\Phi_{ij}$ is the covariance matrix, defined as
\begin{equation} \label{GMFcovariance}
  \Phi_{ij}(\mathbf{x}, \mathbf{x'}) = \langle  \varphi_i(\mathbf{x})  \varphi_j^*(\mathbf{x'}) \rangle,
\end{equation}
where the symbol $\langle \cdots \rangle$ signifies an average over the distribution $P( \bm{\varphi} )$. That is, if $\mathcal{O}(\xbf)$ is some quantity of interest, then 

\begin{equation*}
    \langle  \mathcal{O}(\xbf) \rangle \equiv \int d \bm{\varphi} P(\bm{\varphi}) \mathcal{O}(\xbf).  
\end{equation*}
Notice that the average is taken over field configurations.

In our analysis, we chose not to integrate any prior knowledge about the GMF geometry and statistics, so we use a prior distribution exhibiting statistical isotropy, homogeneity, and mirror symmetry. This is formally encapsulated by writing the Fourier space covariance in the form 
\begin{equation} \label{re_iso-hom}
\langle  \hat{\varphi}_i(\mathbf{k})  \hat{\varphi}^*_j(\mathbf{k}') \rangle = (2 \pi)^3 \delta_{ij}\delta^{(3)}(\mathbf{k}-\mathbf{k}')P(k).
\end{equation}

A crucial point is that the $3$D prior power spectrum $P(k)$ is not known, and is to be inferred as well. It is modeled as a sum of a power law and an integrated Wiener component (\citealt{variable_shadow}). The defining hyperparameters and their prior PDFs (typically called hyperpriors) are summarised in Table \ref{priorparams}, and they are also briefly discussed in \paperI.

\begin{table*}
  \caption[]{Hyperparameters of the prior used in this work}
     \label{priorparams}
     \centering
     \begin{tabular}{c c c c}
        \hline\hline
        Parameter & Distribution & Mean & Standard deviation \\
        \hline
        Total offset ($\mathbf{B_0}$)& Not-applicable & $0$ & Not-applicable  \\
        Total offset st. dev. & Log-normal  & $3$ $\mu$G & $1$ $\mu$G \\
        Total spectral energy  & Log-normal & $1$ $\mu$G & $1$  $\mu$G  \\
        Spectral index  & Normal & $-\frac{11}{3}$ & $1$ \\
        Int. Wiener process amplitude & Log-normal & $1.5$ & $1$ \\ 
        \hline
     \end{tabular}
\end{table*}

\subsection{Sampling the posterior} \label{sec:geoVI}

Equipped with the likelihood and prior, the posterior in terms of the magnetic field $\bm{\mathbf{B}}$ is given by Eq. \ref{Bayes}. Due to the fact that the power spectrum $P(k)$ needs to be inferred along with the configuration of the GMF, this inference problem is non-linear, and cannot be solved by a generalised Wiener filter (\citealt{genWiener}). For this reason, a non-perturbative scheme, called geometrical variational inference (geoVI) developed by \citealt{geoVI} is used.  A brief exposition on geoVI can be found in Appendix A of \paperI. For the purposes of this work it suffices to state that we do not sample magnetic field configurations from the true posterior directly, but rather from an approximate posterior, as is usually the strategy in variational methods. For this task, we employ the Numerical Information Field Theory (\texttt{NIFTy}\footnote{The documentation can be found in \hyperlink{https://ift.pages.mpcdf.de/nifty/index.html}{ift.pages.mpcdf.de/nifty/index.html}.}) package in Python (\citealt{nifty1}; \citealt{nifty3}; \citealt{asclnifty5}, \citealt{niftyre}). The input that is required is the likelihood and the prior of the original physical model, as described in sections \ref{sec:likelihood} and \ref{sec:prior} respectively.

% Finally, we note that the field that serves as a ground truth throughout this work is generated from a dynamo MHD simulation, and in Appendix \ref{sec:groundTruth} the details relevant to the simulation are discussed. The original simulation domain extended to $\sim 1$ kpc in the $x - y$ direction, and $\sim 2$ kpc above the Galactic plane, located roughly at the solar neighbourhood of the Milky Way. 

\subsection{Procedure} \label{sec:gp}

The following is a summary of the specific setting probed in this work and how the synthetic data on which the method is verified is generated. 

\begin{figure}
    \centering
\begin{tikzpicture}[scale=2.5]

    \draw (1,0) arc (0:180:1cm);

    \begin{scope}[viewport={\RotationX}{\RotationY}]

        \draw[variable=\t, smooth] 
            plot[domain=90-\RotationX:-90-\RotationX] (\ToXYZ{\t}{0});
        \draw[densely dashed, variable=\t, smooth]
            plot[domain=90-\RotationX:270-\RotationX] (\ToXYZ{\t}{0});
            
        % \draw[variable=\t, smooth, red] 
        %     plot[domain=90-\RotationY:-90-\RotationY, rotate around y=20] (\ToXYZ{0}{\t});
        % \draw[densely dashed, variable=\t, smooth, red]
        %     plot[domain=90-\RotationY:270-\RotationY, rotate around y=20] (\ToXYZ{0}{\t});
        
        \draw[variable=\t, smooth, purple, thick] 
            plot[domain=85:51, rotate around y=20] (\ToXYZ{0}{\t});

        \draw[variable=\t, smooth, blue, thick] 
            plot[domain=15:70, rotate around y=0] (\ToXYZ{70}{\t});

        \draw[variable=\t, smooth, black, thick] 
            plot[domain=25:80, rotate around x=-40] (\ToXYZ{84}{\t});

        \path (\ToXYZ{15}{55}) node (invisible) {};

        % Label near the invisible node
        \node[above, black] at (invisible) {$\langle \delta \theta\rangle_{\theta |D}$};

        \path (\ToXYZ{75}{35}) node (invisible) {};

        % Label near the invisible node
        \node[above, black] at (invisible) {$\alpha$};

        \path (\ToXYZ{40}{23}) node (invisible) {};

        % Label near the invisible node
        \node[above, black] at (invisible) {$\langle \delta \phi\rangle_{\theta |D}$};

        \node[circle, fill=black, inner sep=1pt, label={210:$\theta_{\text{obs}}$}] at (\ToXYZ{70}{15}) {};
            
        \node[circle, fill=red, inner sep=1pt, label={120:$\theta_{\text{true}}$}, rotate around y=20] at (\ToXYZ{0}{83}) {};

        \path (\ToXYZ{10}{25}) node (invisible) {};

        % Label near the invisible node
        \node[above, black] at (invisible) {$P(\theta |D)$};

          % Parameters for Gaussian distribution
        \def\meanX{20}
        \def\meanY{45}
        \def\stddev{5} % Standard deviation (sqrt of variance)
    
        % Draw 10 random points
        \foreach \n in {1,...,100}{
        \pgfmathsetmacro{\randX}{\meanX + \stddev*rand}
        \pgfmathsetmacro{\randY}{\meanY + \stddev*rand}
        \draw[blue, fill=blue] (\ToXYZ{\randX}{\randY}) circle (0.3pt);
    }
        
    \end{scope}

\end{tikzpicture}

\caption{Illustration of relevant angles on the sky. A UHECR of known rigidity $r_*$ enters the Galaxy with an arrival direction $\theta_\text{true}$ (red dot). Because of the GMF, it is deflected and is observed on Earth as arriving from $\theta_\text{obs}$ (black dot). The angular distance between $\theta_\text{obs}$ and $\theta_\text{true}$ is $\alpha$, and it is the error that the GMF induces on the observed arrival direction. We backtrack the particle through each GMF configuration sampled using \texttt{NIFTy}, thus ending up with a distribution of arrival directions $P(\theta |D)$, with $D$ defined in Eq. \ref{eq:D}. From the posterior samples drawn, we calculate the mean angular distances $\langle \delta \theta \rangle_{\theta | D}$ and $\langle \delta \phi \rangle_{\theta | D}$ to the true and observed arrival directions, respectively, as well as the standard deviations for the former. Note that the scales in this artificial example are exaggerated for visual clarity, and do not correspond to an application of the method.}
\label{fig:POS-explanation}
\end{figure}
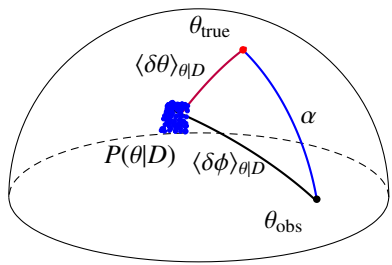

\begin{itemize}
    \item \textbf{Spatial domain:} The modeled space is assumed to be periodic due to implementation details of the ground truth, and also we pad our space by a factor of two, and so the $x$ and $y$ directions reach an extent of $\sim 2$ kpc. The resulting cube is partitioned uniformly into $N_x$,  $N_y$, and  $N_z$ segments per axis, where $N_x = N_y = 48$, and $N_z = 64$, with padding. In that setting, every voxel has a linear dimension of approximately $30$ pc. This can accomodate the expected size of the dust clouds, as well as the uncertainty of the measurement's positions - at least as an order of magnitude (\citealt{Pelgrims}).
    
    \item \textbf{Data masking:} We apply $R_\text{local}$ (see section \ref{sec:likelihood}) to the ground truth field, in order to acquire the noiseless data.  

     \item \textbf{Adding noise to local data:} Gaussian noise with covariance matrix $N_\text{local}$ ( Eq. \ref{noise_cov_local}) is added to each observed data vector. 
     
    \item \textbf{Integrated data}: Optionally (see section \ref{section:faraday}), the likelihood is supplemented by an additional term for the integrated local measurements, as in Eq. \ref{eq:likelihood}. In practice, the magnetic field is transformed from a Cartesian coordinate system to a spherical polar coordinate system with the Earth at the origin. Then, the radial component of the GMF - which is equivalent to the LOS component - is integrated along individual LOSs, resulting in a set of $2$D integrated measurements that inform the model further.

    \item \textbf{Adding noise to integrated data:} Gaussian noise with covariance $N_\text{int}$ (Eq. \ref{noise_cov_int}) is added to each pixel on the celestial sphere, to contaminate the data acquired from the previous step.
    
    \item \textbf{Sampling the approximated posterior:} Finally, the geoVI method is applied to the true posterior distribution, resulting in samples from the approximate distribution. To all the latent fields sampled, the projection operator (Eq. \ref{projector}) is applied once more, in order to obtain posterior samples of the divergence-free GMF.

    \item \textbf{Application to UHECR backtracking:} Through each of the GMF samples drawn from \eqref{Bayes} in the previous step, we backtrack a UHECR of known observed arrival direction $\theta_\text{obs}$ and rigidity $r_* \equiv E/Z$. Recording the final velocity of the particles, in particular their original directions $\theta$ when they leave $\mathcal{V}$, essentially provides samples from the distribution $P(\theta|D)$ of the particles' original arrival directions before entering the GMF, conditional to the data

    \begin{equation} \label{eq:D}
        D \equiv \{d, r_*, \theta_\text{obs} \}.
    \end{equation}

    To keep the discussion simple, in this work we only consider UHECRs of fixed rigidity $r_* = 5 \times 10^{19}$ eV (equivalently, protons of energy $E= 5 \times 10^{19}$ eV). As a way to benchmark the quality of our reconstructions in the context of UHECR physics, we will compare the angular separation $\delta \theta$ between the true arrival direction $\theta_\text{true}$ and that of the backpropagated UHECR, ending up with a distribution over $\delta \theta$. In this context, the `true arrival direction' always refers to the UHECR's direction right where it entered $\mathcal{V}$. In Fig. \ref{fig:POS-explanation}, we provide a visual representation of the quantities defined in this section.

\end{itemize}

\section{Results}\label{sec:results}

In this section, we use \texttt{NIFTy} in order to sample the posterior distribution for three different scenarios: In scenario A, the observed data consist of local measurements only, and at each location only the components of the GMF that are parallel to the POS are probed.  In scenario B, all three components of the GMF (including the LOS) are probed on equal footing, for comparison. Finally, in scenario C, we use the same dataset as in scenario A, but additionally use integrated LOS information over the whole sky. 

For each of these scenarios, we will benchmark the success of the reconstruction by using it in order to infer the true arrival direction of a UHECR with fiducial rigidity of $r_* = 5 \times 10^{19}$ eV for all possible observed arrival directions on the northern sky, as described in the previous section. 

\subsection{Scenario A: Local measurements with POS information only} \label{sec:no_los}

The local GMF information that one can acquire through starlight polarization-based tomography alone is confined to the celestial sphere (\citealt{Panopoulou17}; \citealt{Pelgrims}). For that reason, in this section, we will sample the posterior Eq. \ref{Bayes} conditional to local GMF data $d$ that are completely blind to the LOS dimension, as is the case for polarization measurements. 

To that end, we will work on a spherical polar coordinate system with the Sun at the origin. The magnetic field is expressed as $\mathbf{B}(\mathbf{x}) = (B_r, B_\theta, B\phi)$ in that coordinate system. In Fig. \ref{fig:ex1} we perform the reconstruction of the simulated GMF described in Appendix \ref{sec:groundTruth}. In Fig. \ref{fig:1a} the ground truth is shown. Fig. \ref{fig:1b} depicts the synthetic local GMF data obtained from the ground truth for this scenario. The result of the reconstruction algorithm is a set of $100$ posterior samples of Eq. \ref{Bayes}, given the data of Fig. \ref{fig:1b}. In Fig. \ref{fig:1c}, the mean of the posterior samples is shown. 

In Figs. \ref{fig:meana} and \ref{fig:stda} we show the mean and standard deviation of the angular distance error ($\langle \delta \theta\rangle_{\theta |D}$ and $\sigma_{\theta |D}$ respectively) obtained through the use of the GMF reconstructions shown in Fig. \ref{fig:ex1}. Observe that $\langle \delta \theta\rangle_{\theta |D}$ and $\sigma_{\theta |D}$ vary across the celestial sphere, and the specific structure of these functions depends on the specific ground truth GMF chosen. That said, the greatest error of the reconstruction for this setting is approximately $14^\circ$. In order to judge the performance, in Fig. \ref{fig:noRecon}  we depict the angular error in the arrival direction assuming the observed ones were true - that is, ignoring the correction using the recovered GMF. Comparing Fig. \ref{fig:noRecon} to Fig. \ref{fig:meana}, we observe that reconstructing the local GMF conditional to $d$ yields a significant improvement in our ability to recover UHECR arrival directions. This result suggests that $\langle \delta \theta\rangle_{\theta |D}$ is greater for UHECRs observed to arrive from directions where the influence of the GMF is greater (Fig. \ref{fig:noRecon}), in this case at small longitudes. This correlation will be explored further in section \ref{sec:bias}.

\begin{figure*}
    \centering
    \begin{subfigure}{.45\textwidth}
        \centering
        \includegraphics[scale=0.15]{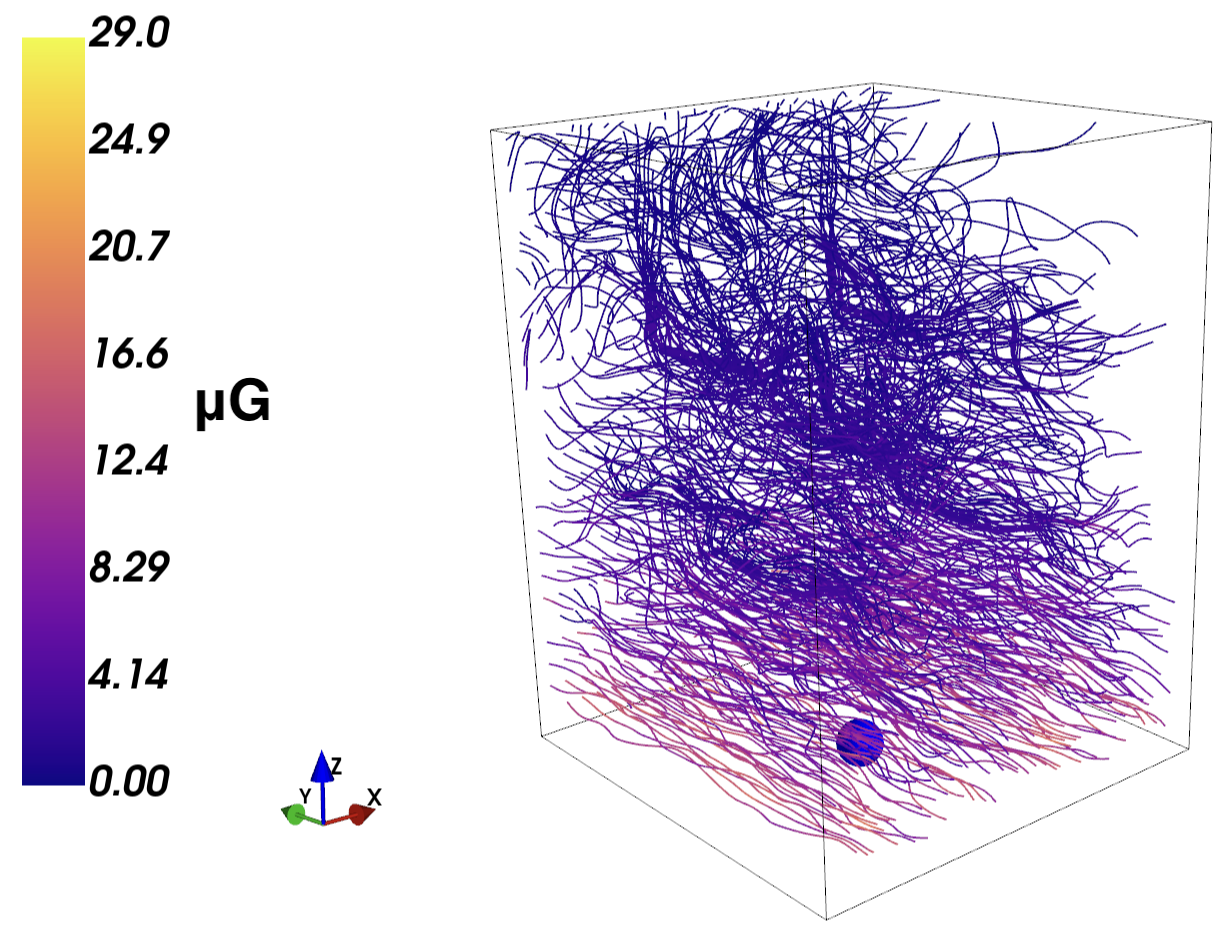}
        \caption{The ground truth.}
        \label{fig:1a}
    \end{subfigure}
    \hfill
    \begin{subfigure}{.45\textwidth}
        \centering
        \includegraphics[scale=0.15]{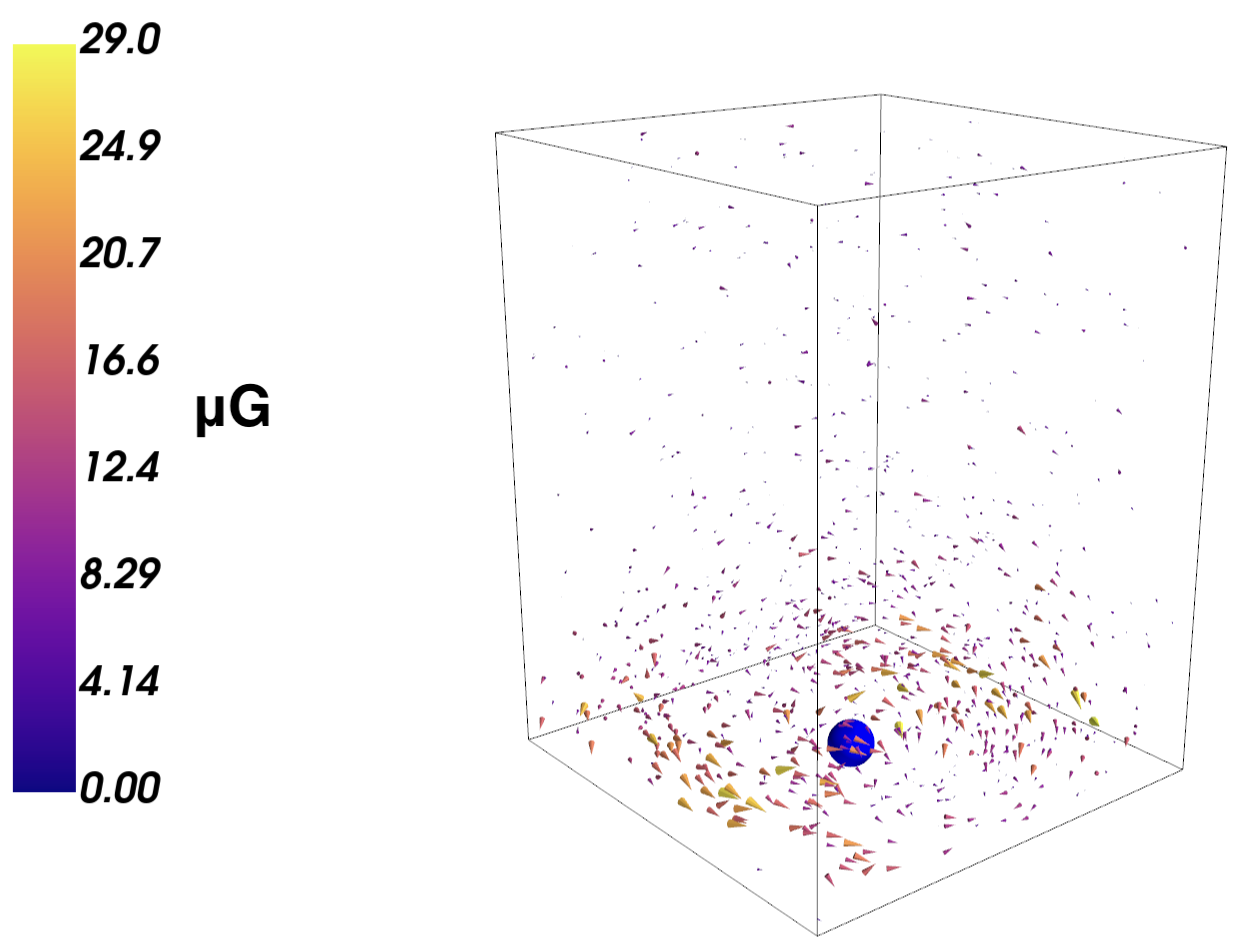}
        \caption{The local and sparse data, confined to the POS.}
        \label{fig:1b}
    \end{subfigure}
    \vspace{0.5cm} % Adds vertical space between the rows
    \begin{subfigure}{.45\textwidth}
        \centering
        \includegraphics[scale=0.15]{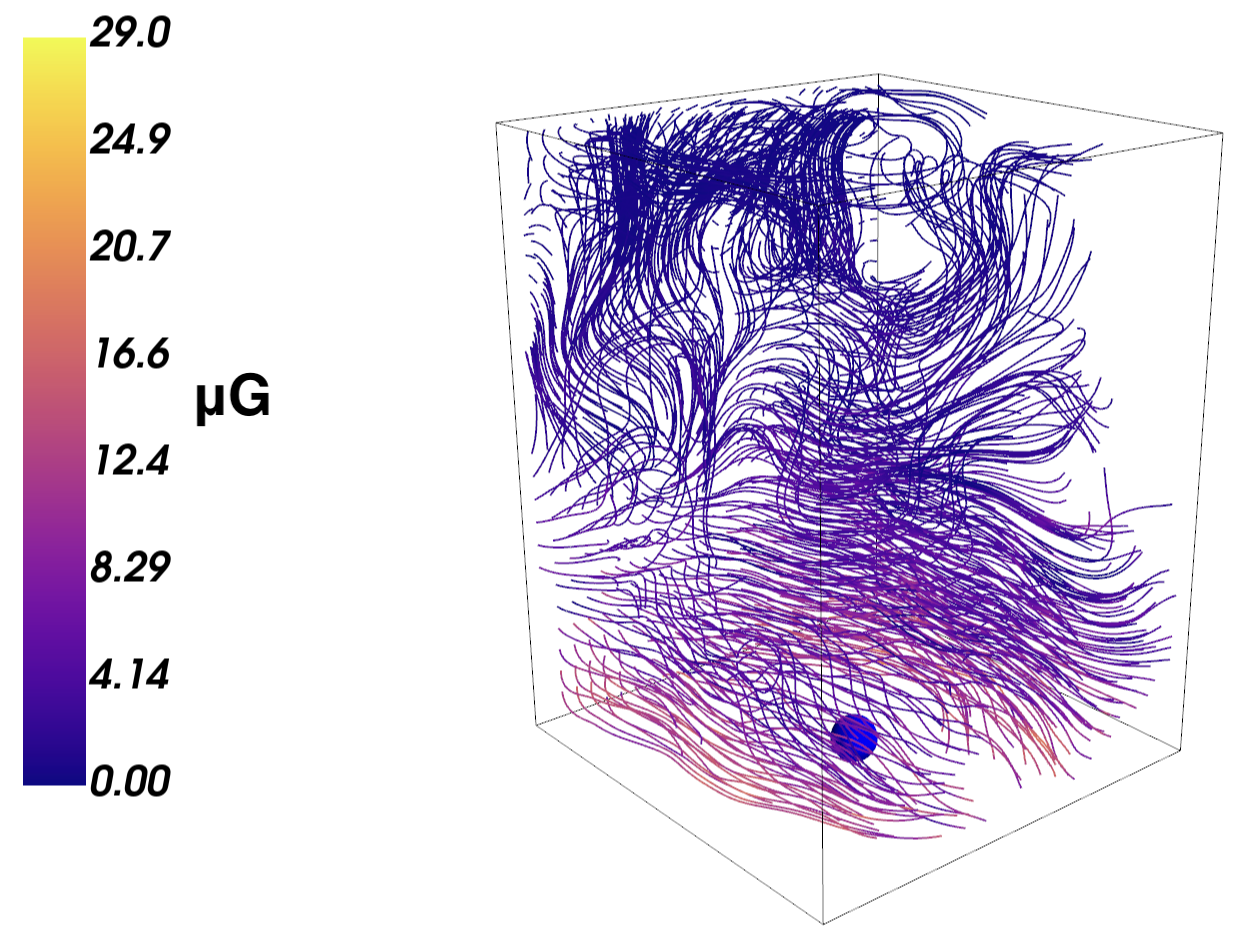}
        \caption{Mean of the posterior distribution conditional to the data of Fig. \ref{fig:1b}.}
        \label{fig:1c}
    \end{subfigure}
    \hfill
    \begin{subfigure}{.45\textwidth}
        \centering
        \includegraphics[scale=0.22]{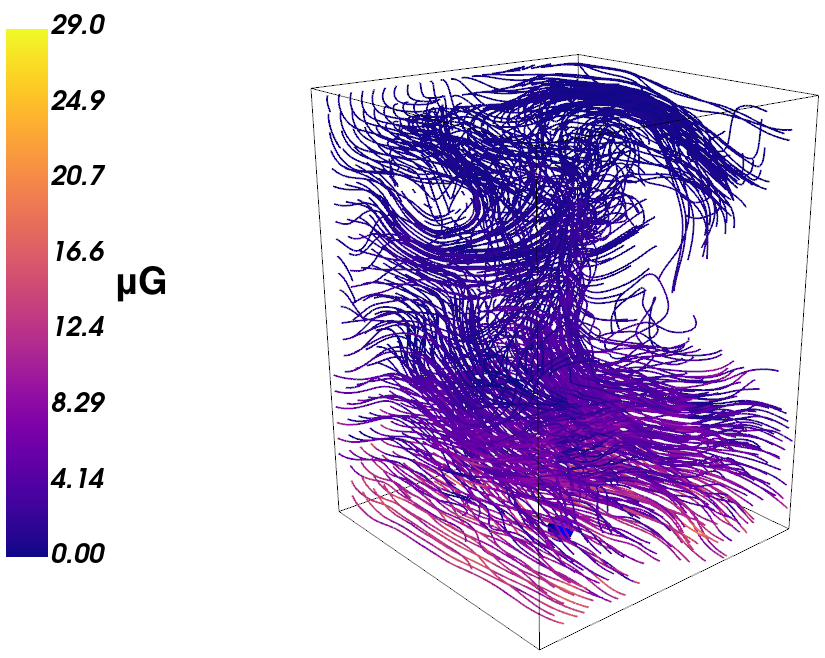}
        \caption{Mean of the posterior distribution conditional to the data of Figs \ref{fig:1b} and \ref{fig:faradayb}.}
        \label{fig:1d}
    \end{subfigure}
    \caption{Reconstruction of the simulated 3D magnetic field with the use of local data that lack LOS field component information. The blue sphere represents the celestial sphere. \textbf{Top Left:} The ground truth; the GMF obtained as described in Appendix \ref{sec:groundTruth}. The field is rescaled so that it has a RMS norm of $5$ $ \mu$G. \textbf{Top Right:} Synthetic data based on the ground truth of Fig. \ref{fig:1a}. Note that the radial component of the magnetic field is not measured. \textbf{Bottom Left}: The mean of the approximating posterior distribution attained via the geoVI algorithm based on the data provided in Fig. \ref{fig:1b}.\textbf{Bottom Right}: The mean of the approximating posterior distribution attained conditional to the local data of Fig. \ref{fig:1b} as well as integrated measurements of the radial component (Fig. \ref{fig:faradayb}).}
    \label{fig:ex1}
\end{figure*}

\begin{figure}
        \centering
        \begin{subfigure}{.35\textwidth}
         \centering
         \includegraphics[scale = 0.35]{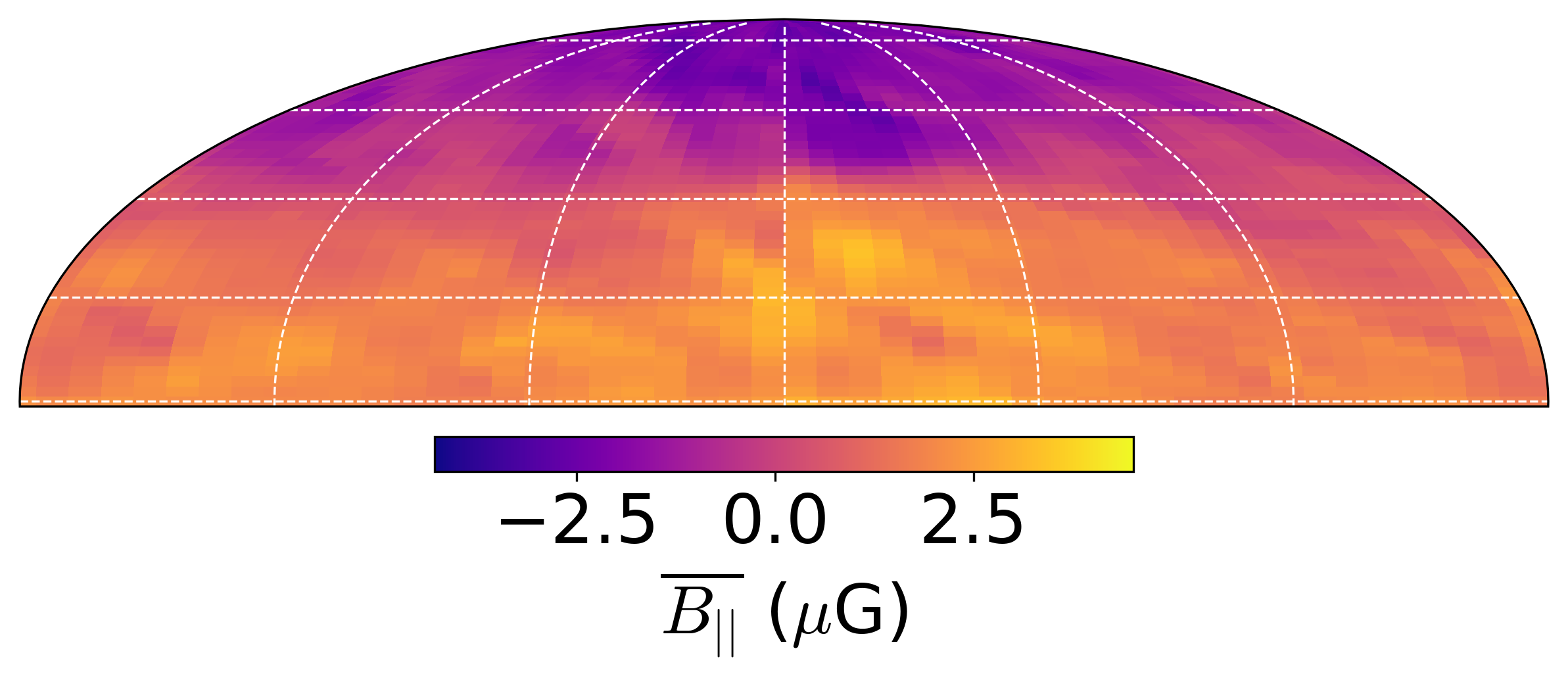}
         \caption{The integrated LOS component of the ground truth field, shown in Fig. \ref{fig:1a}.}
         \label{fig:faradaya}
     \end{subfigure}
     \hfill
     \begin{subfigure}{.35\textwidth}
         \centering
         \includegraphics[scale = 0.35]{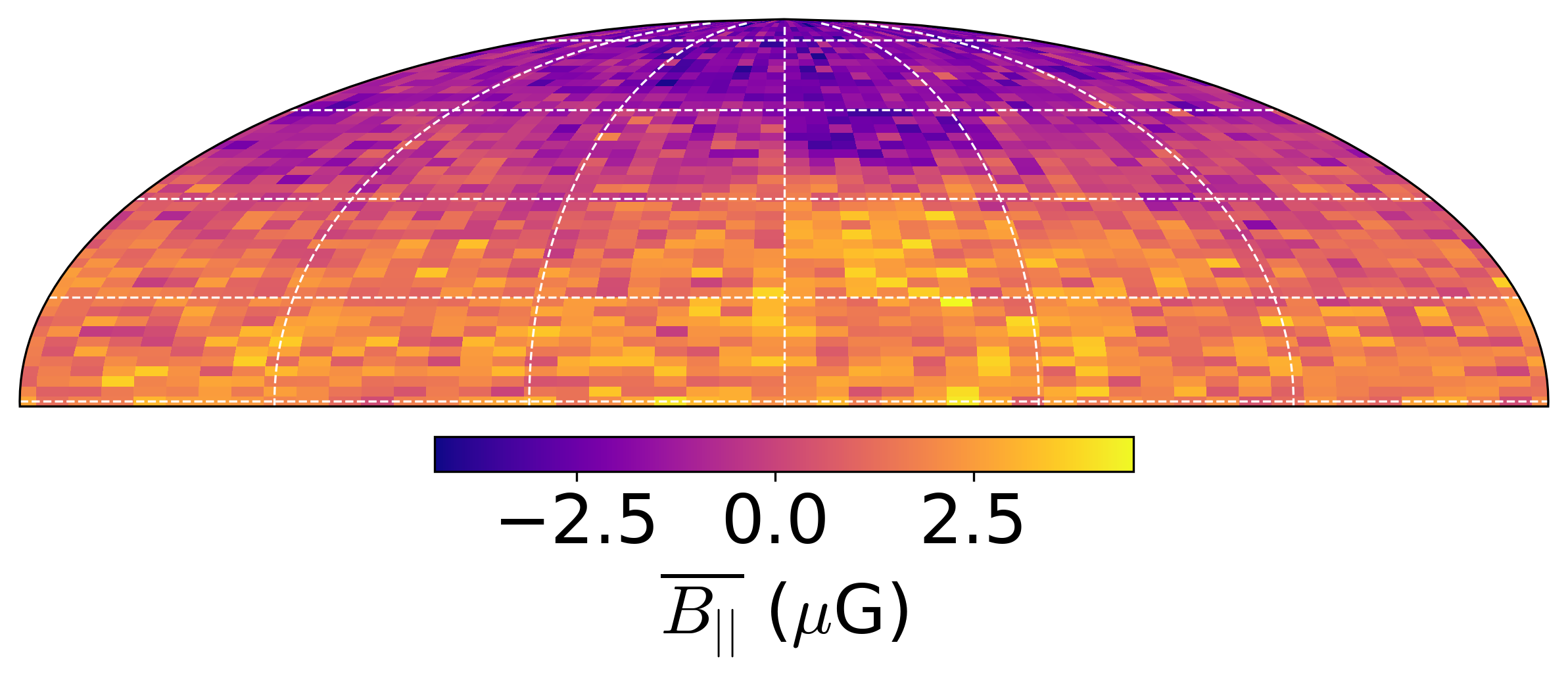}
         \caption{As in Fig. \ref{fig:faradaya}, with Gaussian noise contamination.}
         \label{fig:faradayb}
     \end{subfigure}
     \hfill
     \begin{subfigure}{.35\textwidth}
         \centering
         \includegraphics[scale = 0.35]{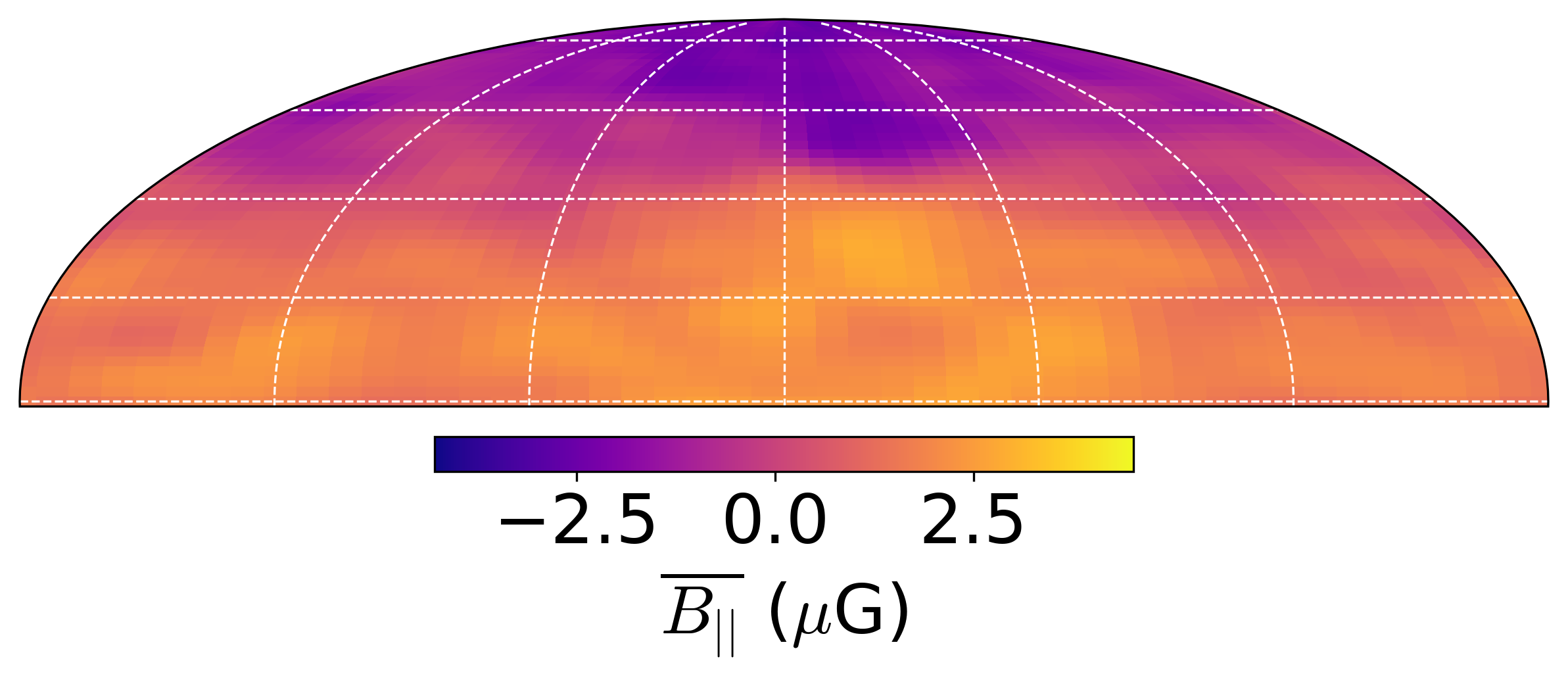}
         \caption{The integrated LOS component of the posterior mean, conditional to the data of Figs. \ref{fig:1b} and Fig. \ref{fig:faradayb}.}
         \label{fig:faradayc}
     \end{subfigure}
     \caption{ \textbf{Top}: Averaged LOS component of the test magnetic field, shown in Fig. \ref{fig:1a}. \textbf{Middle}: Noisy integrated data that is used along with the sparse and local data shown in Fig. \ref{fig:1b} in order to define the LOS-informed posterior distribution. The noise covariance is set to $0.5$ $\mu$G$^2$, while the density of integrated measurements is $0.1$ deg$^{-2}$  \textbf{Bottom}: Averaged LOS component of the mean $3$D configuration of the approximating posterior distribution conditional to the data of Figs. \ref{fig:1b} and \ref{fig:faradayb}. 
        }
        \label{fig:faraday}
    \end{figure}

\begin{figure*}
    \centering
    % First row with figures (a) and (b)
    \begin{minipage}{.5\textwidth}
        \centering
        \begin{subfigure}{\linewidth}
            \includegraphics[width=\linewidth]{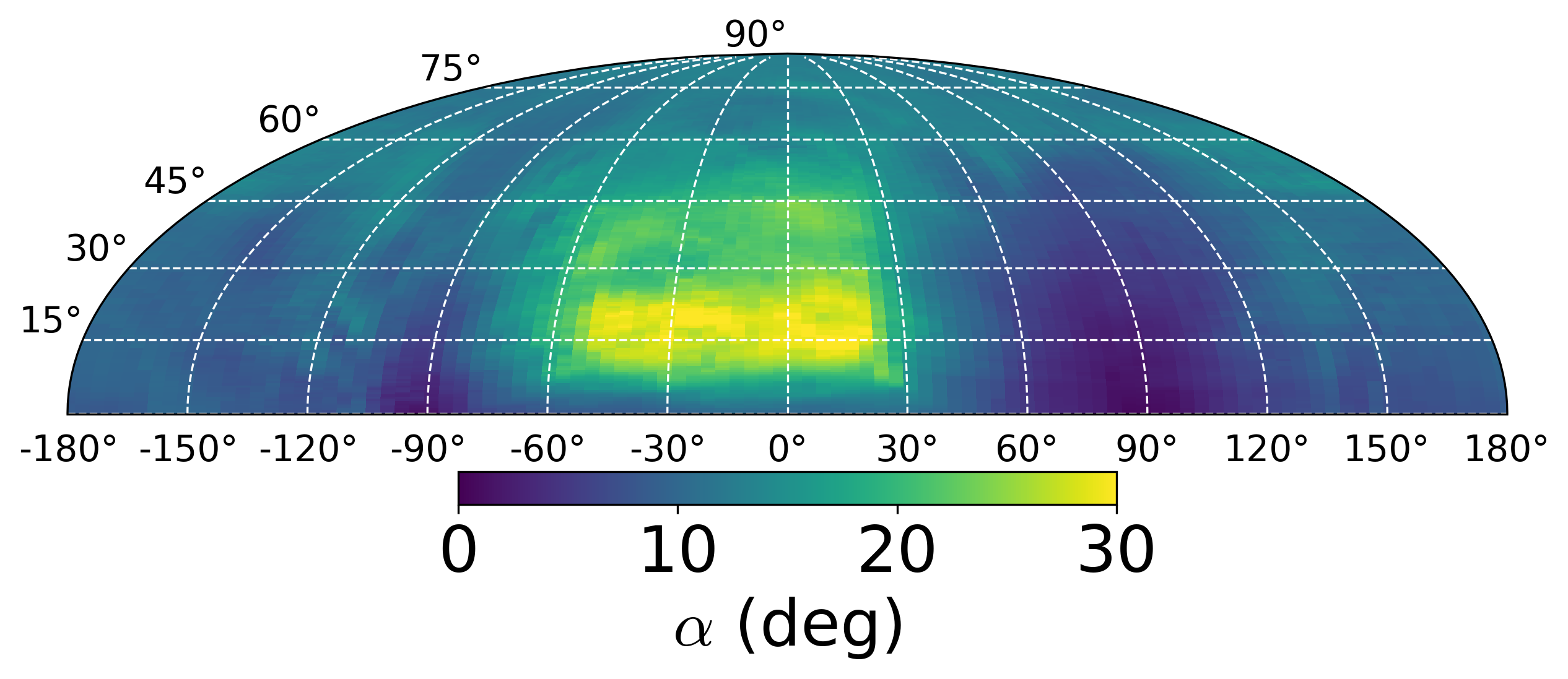}
            \caption{Deflection map for the ground truth.}
            \label{fig:noRecon}
        \end{subfigure}
    \end{minipage}%
    \begin{minipage}{.5\textwidth}
        \centering
        \begin{subfigure}{\linewidth}
            \includegraphics[width=\linewidth]{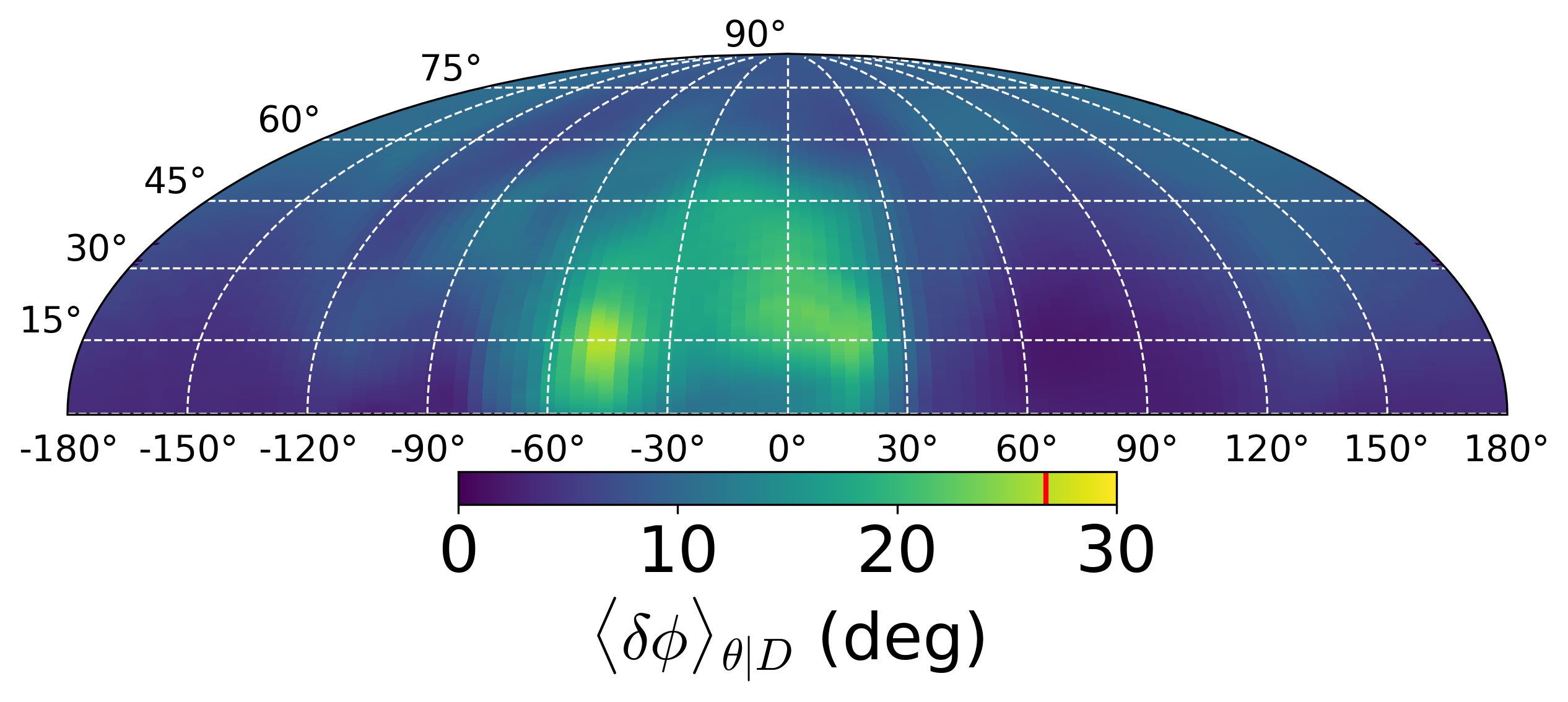}
            \caption{Mean deflection for scenario A.}
            \label{fig:deflectionMapFFF}
        \end{subfigure}
    \end{minipage}
    % Second row with figures (c) and (d)
    \begin{minipage}{.5\textwidth}
        \centering
        \begin{subfigure}{\linewidth}
            \includegraphics[width=\linewidth]{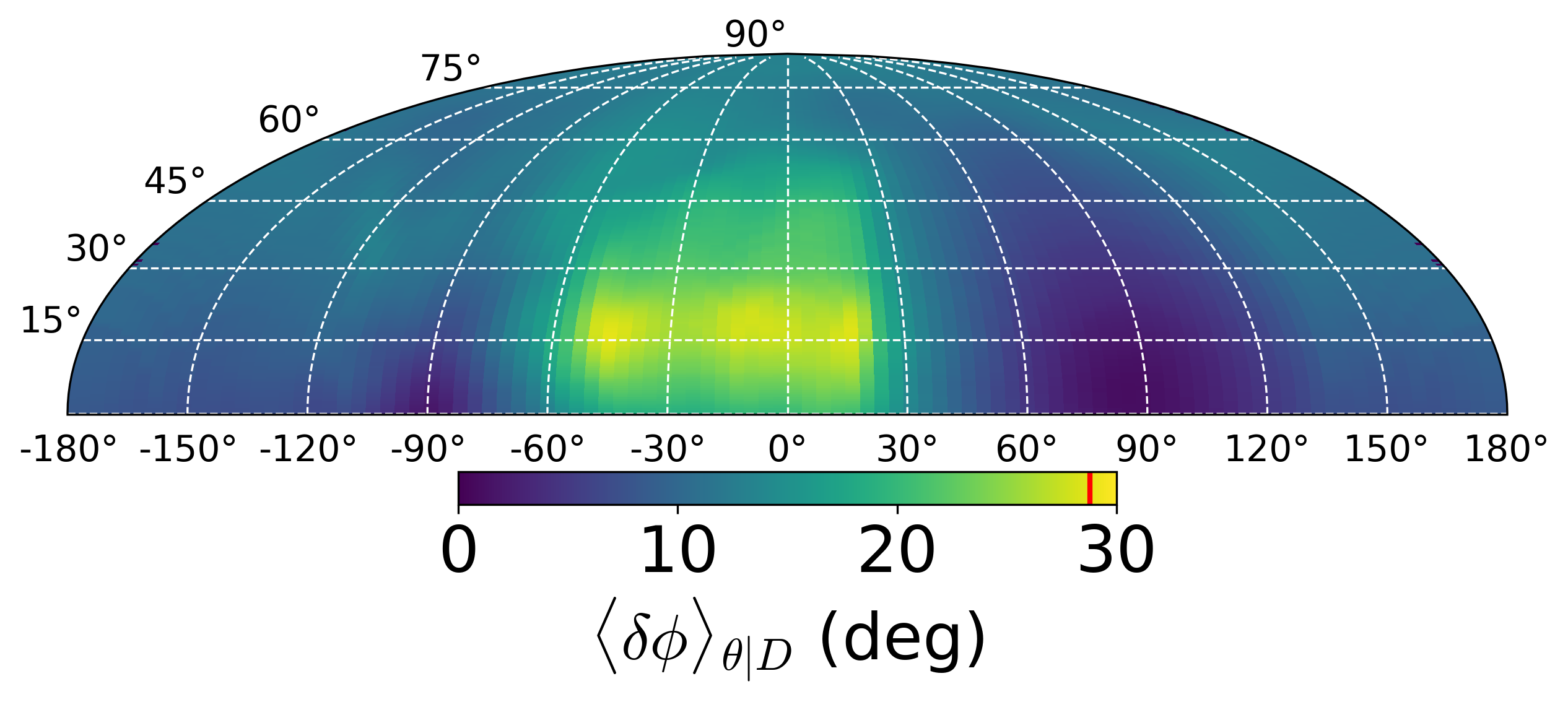}
            \caption{Mean deflection for scenario B.}
            \label{fig:deflectionMapFTF}
        \end{subfigure}
    \end{minipage}%
    \begin{minipage}{.5\textwidth}
        \centering
        \begin{subfigure}{\linewidth}   \includegraphics[width=\linewidth]{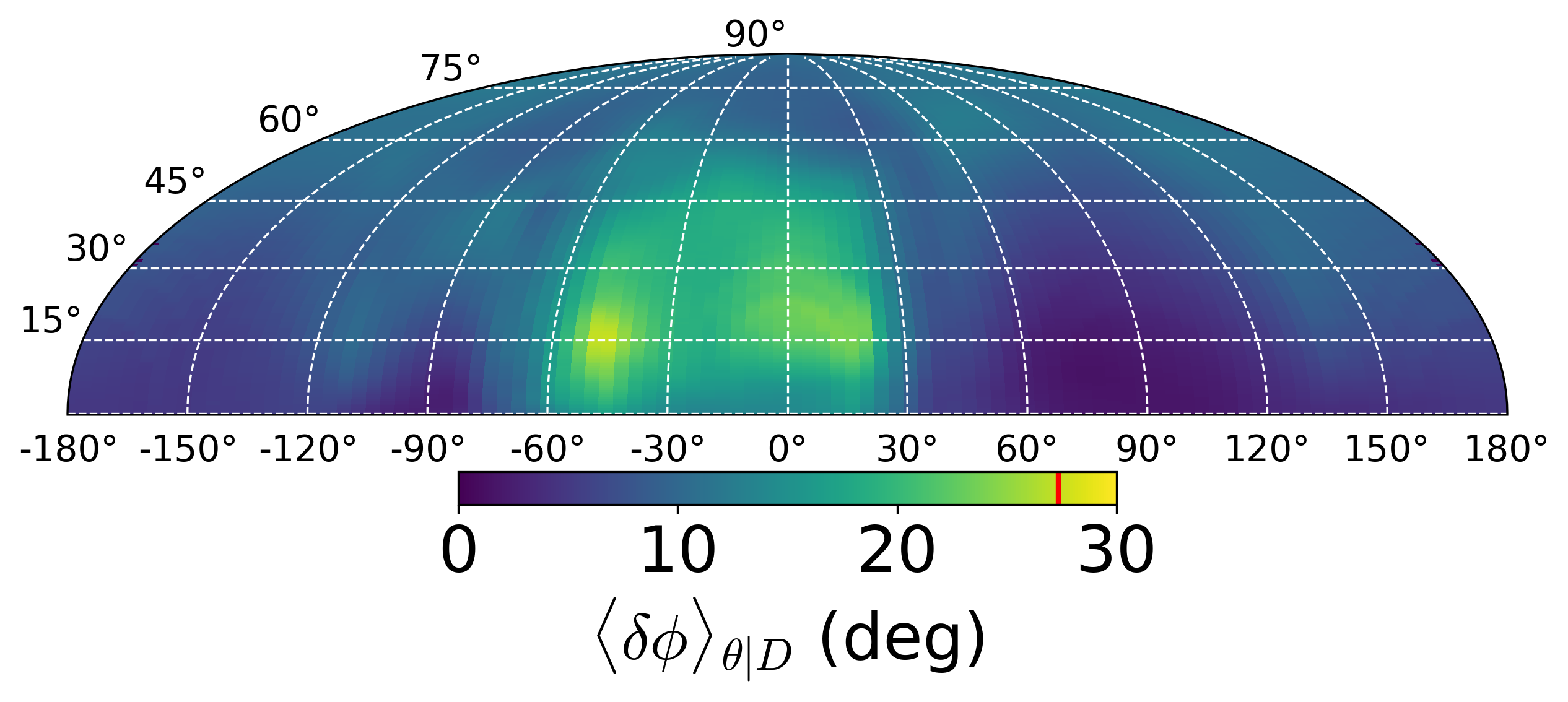}
            \caption{Mean deflection for scenario C.}
            \label{fig:deflectionMapFFT}
        \end{subfigure}
    \end{minipage}
    \caption{Amount by which a UHECR of rigidity $r_* = 5 \times 10^{19}$ eV is deflected by different GMF configurations as a function of its observed arrival direction on Earth (the deflection map - see Fig. \ref{fig:POS-explanation} for the definition of the relevant angles). \textbf{Top left}: True deflection map. \textbf{Top right}:  The mean deflection over the posterior samples for scenario A. \textbf{Bottom left}: As in \ref{fig:deflectionMapFFF}, but the local measurements of the GMF now contain information on the LOS component as well as the POS component (scenario B).  The additional information in this case causes a greater resemblance of the posterior mean to the true field, and so the deflection map is closer to Fig. \ref{fig:noRecon}. \textbf{Bottom right}: As in \ref{fig:deflectionMapFFF}, but the posterior is additionally constrained by the integrated data seen in Fig. \ref{fig:faradayb} (scenario C). The colobar scale is kept up to $30$ degrees to aid visual comparison. The red line on the colorbar indicates the maximum deflection for each case. Notice that the dominant central feature of Fig. \ref{fig:noRecon} is recovered in Figs. \ref{fig:deflectionMapFFF} - \ref{fig:deflectionMapFFT}, since it is caused by the largest scale features of the magnetic field, which we are able to infer in every case.}
    \label{fig:bias}
\end{figure*}

\begin{figure*}
    \centering
    \begin{minipage}{.5\textwidth}
        \centering
        \begin{subfigure}{\linewidth}
            \includegraphics[width=\linewidth]{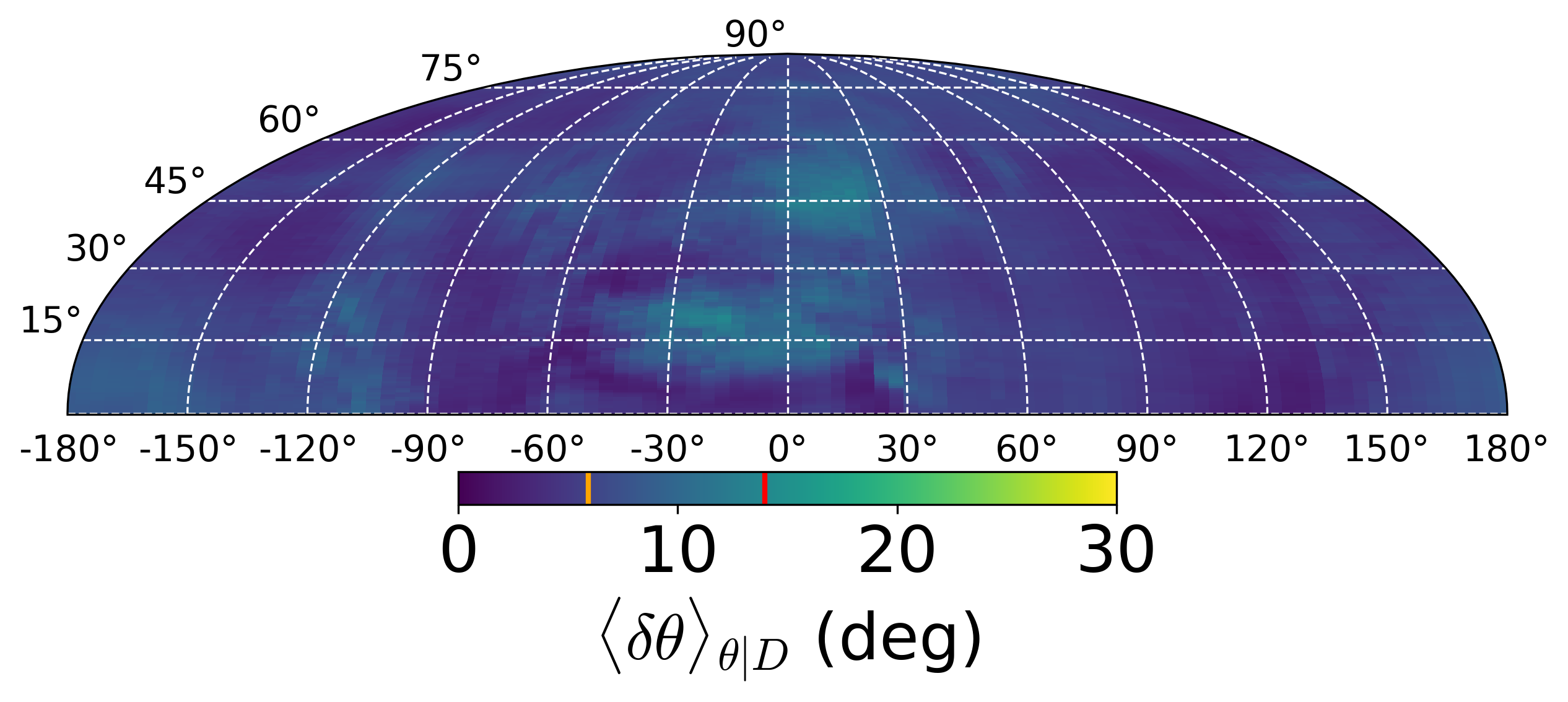}
            \caption{Scenario A}
            \label{fig:meana}
        \end{subfigure}%
    \end{minipage}%
    \hfill
    \begin{minipage}{.5\textwidth}
        \centering
        \begin{subfigure}{\linewidth}
            \includegraphics[width=\linewidth]{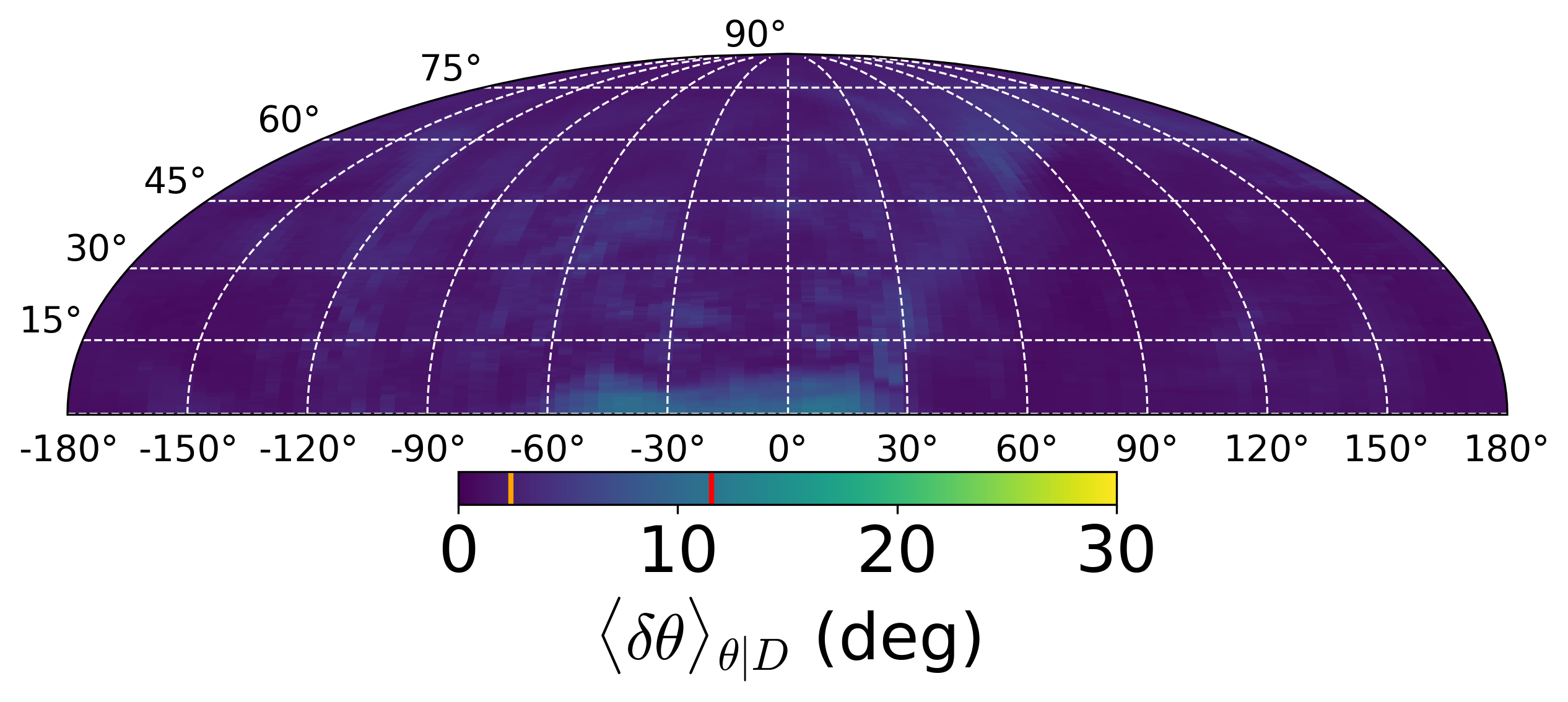}
            \caption{Scenario B}
            \label{fig:meanb}
        \end{subfigure}
    \end{minipage}
    
    \begin{minipage}{\textwidth}
        \centering
        \begin{subfigure}{0.5\linewidth}
            \includegraphics[width=\linewidth]{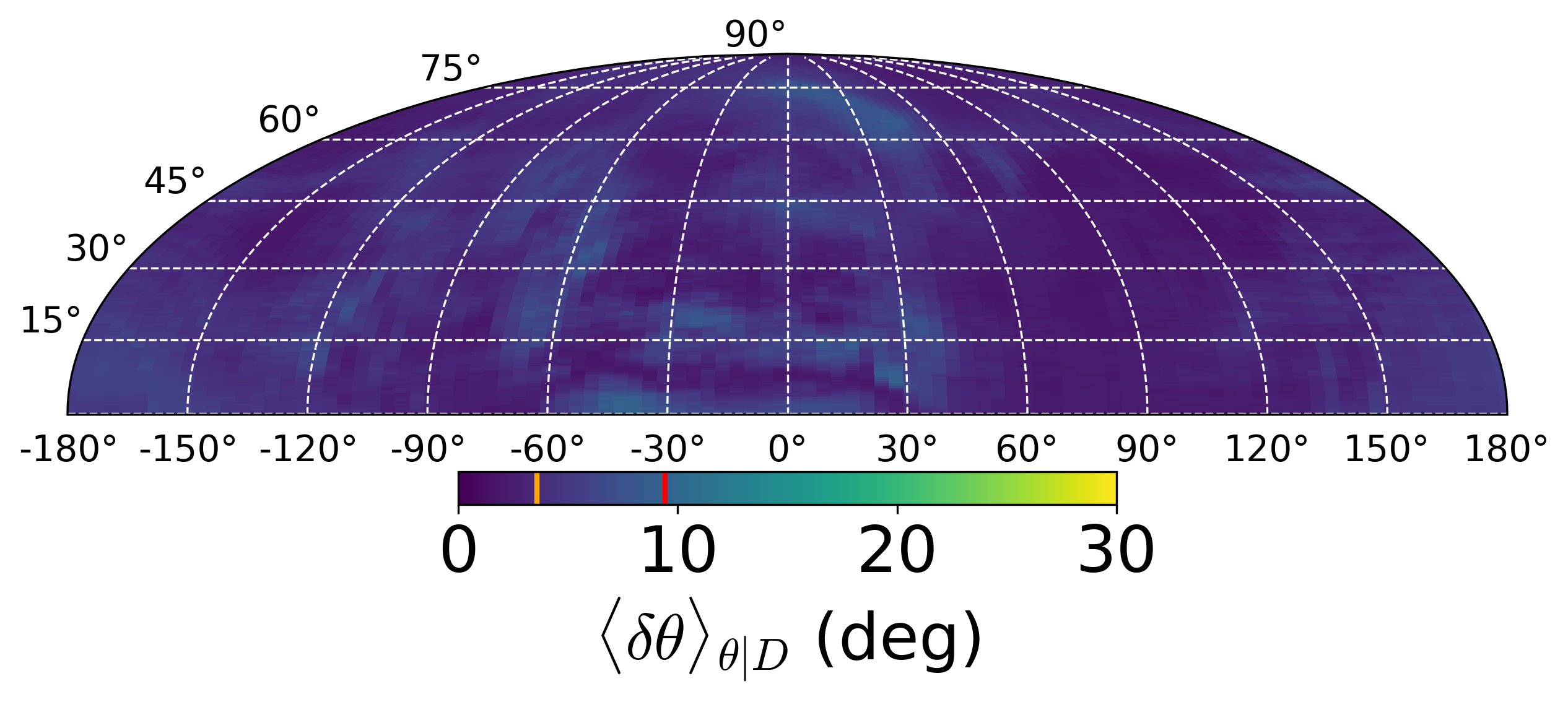}
            \caption{Scenario C}
            \label{fig:meanc}
        \end{subfigure}
    \end{minipage}
    \caption{Mean angular error of the reconstruction (see Fig. \ref{fig:POS-explanation}) as a function of all possible arrival directions on the Northern hemisphere, for the case of a UHECR of rigidity $r_* = 5 \times 10^{19}$ eV. \textbf{Top left}: The magnetic field data consist of local information with the LoS component is projected out (scenario A). \textbf{Top right}: The magnetic field data consist of local information with the LOS component measured (scenario B) \textbf{Bottom}: As in top left, but the data is supplemented by integrated LOS data (scenario C)(see Fig. \ref{fig:faraday}). The colorbar scale is kept up to $30$ degrees to aid visual comparison with Fig. \ref{fig:noRecon}. The red and orange lines on the colorbar indicate the maximum and mean values of the map, respectively.}
    \label{fig:mean}
\end{figure*}

\begin{figure*}
    \centering
    \begin{minipage}{.5\textwidth}
        \centering
        \begin{subfigure}{\linewidth}
            \includegraphics[width=\linewidth]{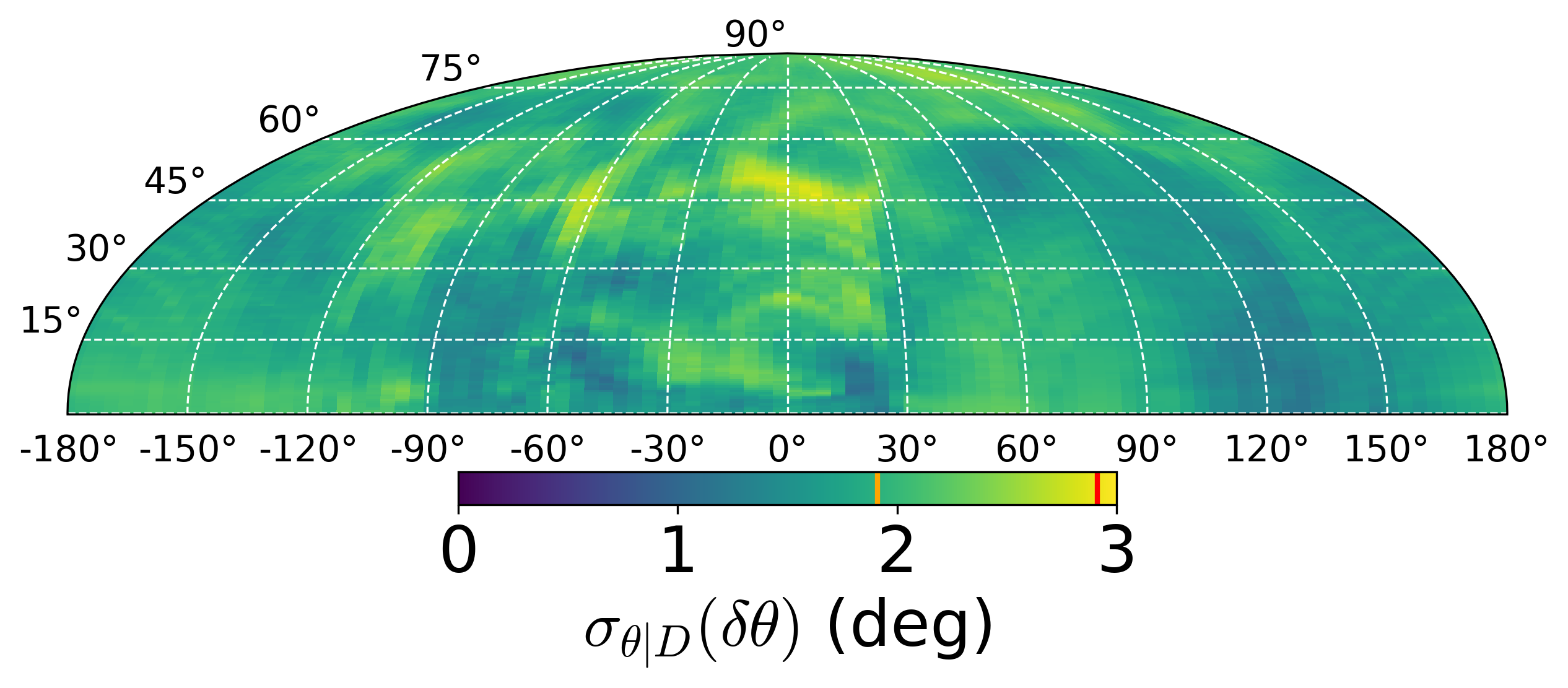}
            \caption{Scenario A}
            \label{fig:stda}
        \end{subfigure}%
    \end{minipage}%
    \hfill
    \begin{minipage}{.5\textwidth}
        \centering
        \begin{subfigure}{\linewidth}
            \includegraphics[width=\linewidth]{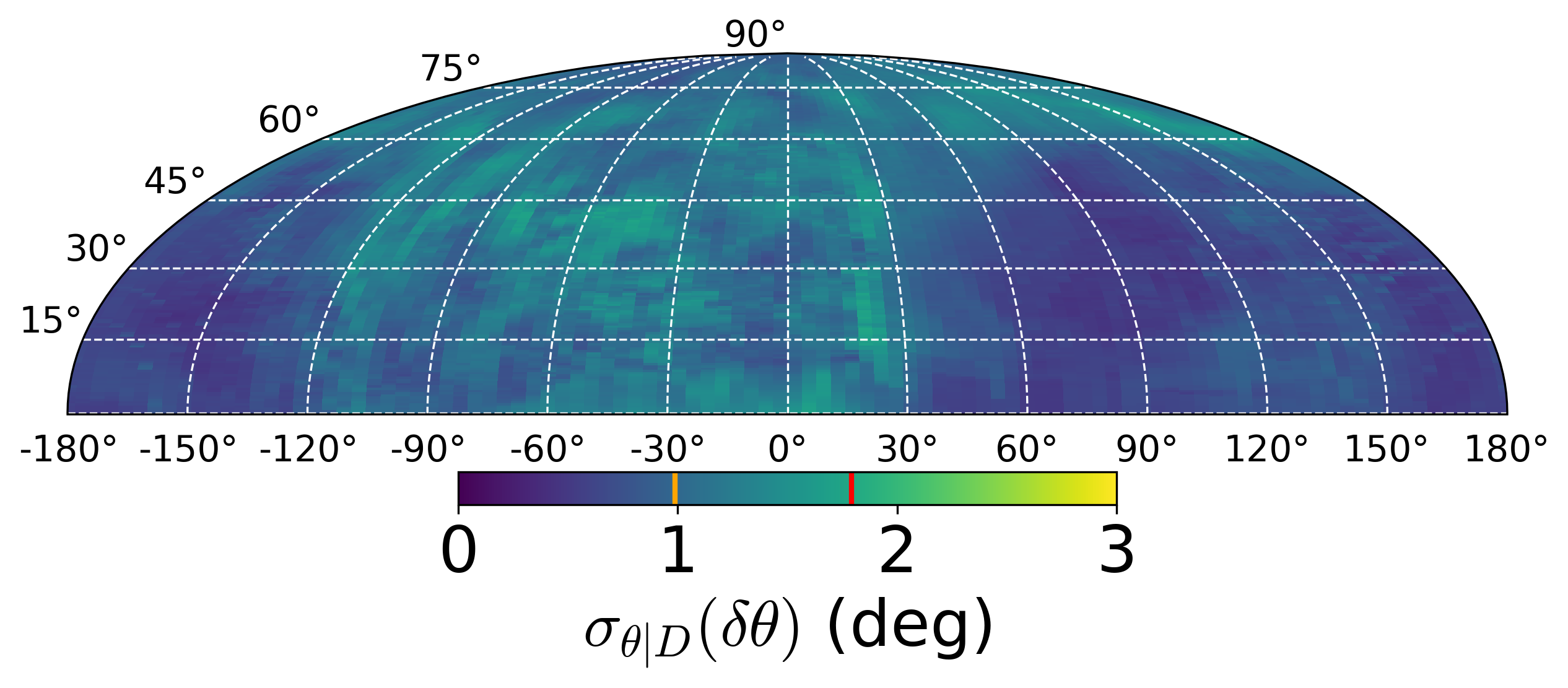}
            \caption{Scenario B}
            \label{fig:stdb}
        \end{subfigure}
    \end{minipage}
    
    \begin{minipage}{\textwidth}
        \centering
        \begin{subfigure}{0.5\linewidth}
            \includegraphics[width=\linewidth]{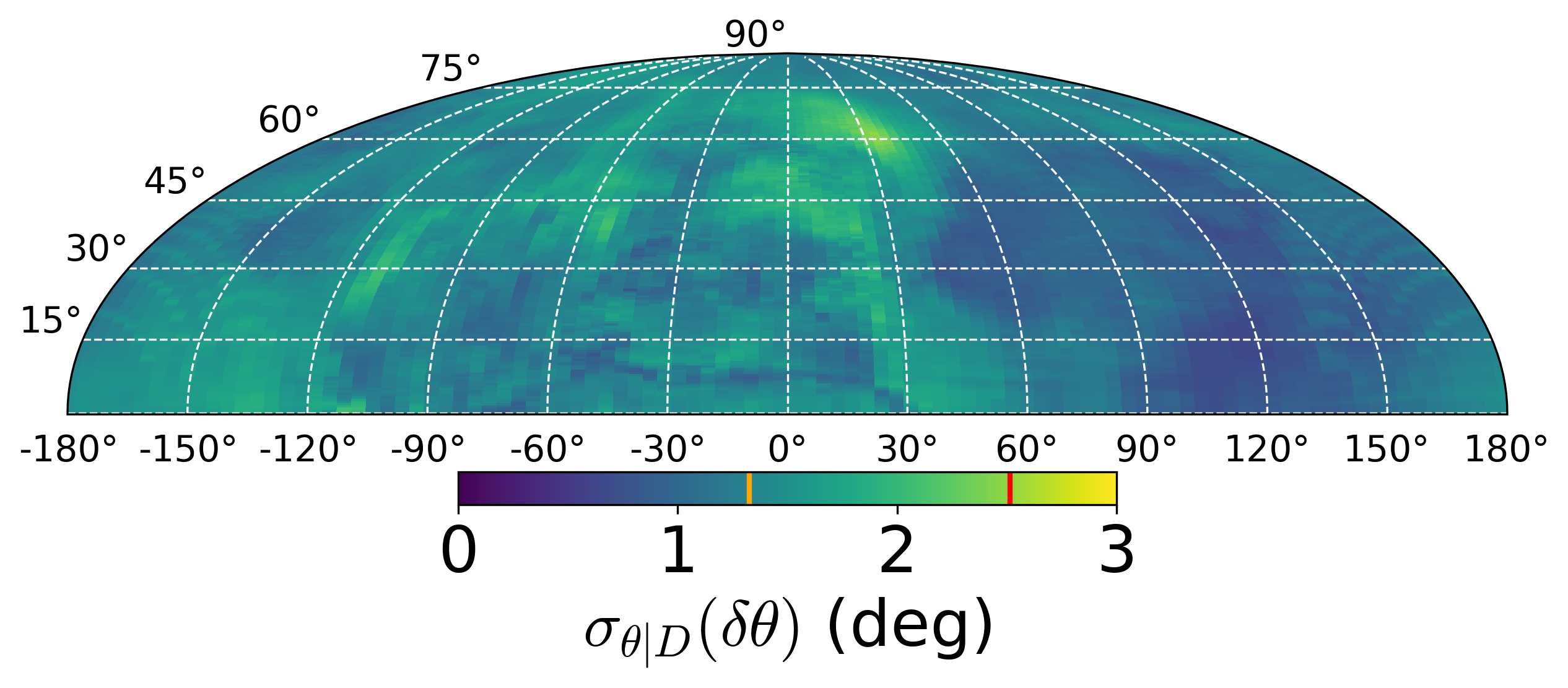}
            \caption{Scenario C}
            \label{fig:stdc}
        \end{subfigure}
    \end{minipage}
    \caption{As in Fig. \ref{fig:mean}, but for the corresponding angular error standard deviations as a function of observed arrival direction.}
    \label{fig:std}
\end{figure*}

\subsection{Scenario B: Local measurements with full $3$D information at each measured location}  \label{sec:full_3d}

In this section we examine the impact that a complete lack of observation of the LOS (scenario A) has on the UHECR arrival direction reconstruction. For that purpose, we perform the same inference as in section \ref{sec:no_los} with the difference that now the LOS component is also probed locally, just like the POS components. In Figs. \ref{fig:meanb} and \ref{fig:stdb} we plot the mean angular error $\langle \delta \theta \rangle_{\theta | D}$ and the respective standard deviation, for this scenario. Comparing to the results of scenario A (see Figs. \ref{fig:meana} and \ref{fig:stda}), we observe that the quality of the reconstruction greatly improves when local LOS information is included. While the maximum-occurring mean angular error drops by a few degrees, in general the improvement is dramatic in that the total area of the sky where the maximum bias occurs is substantially reduced. This observation also holds for the variance.

While we consider $\theta_\text{obs}$ over the whole northern hemisphere for benchmarking purposes, in real applications only sufficiently high Galactic latitudes are relevant. The reason for this is that we aim to reconstruct the GMF at a scale of up to a couple of kpcs, and therefore we must choose UHECRs that have traveled through the part of the Galaxy whose GMF we reconstruct. That said, especially at the physically relevant case of high Galactic latitudes, the inclusion of local LOS information dramatically improves the backtracking results. 

We have shown that knowledge of local LOS information would yield a substantial improvement over our ability to reconstruct the GMF, at least as far as UHECR backtracking is concerned. As stellar polarization data alone cannot probe the LOS dimension, this information would have to be supplemented by additional methods (e.g. Zeeman measurements). However, measurement of the LOS GMF component locally is a notoriously difficult task, and so in what follows, we will attempt to mitigate this by including integrated LOS information in our likelihood. 

\subsection{Scenario C: Local measurements with POS information supplemented by integrated LOS measurements for the whole sky} \label{section:faraday}

In this section we consider the inclusion of integrated constraints on the LOS component of the GMF as shown in Fig. \ref{fig:faradayb}, while the local measurements at the dust clouds, simulating those obtained through polarised starlight tomography, are still projected on the celestial sphere as in Fig. \ref{fig:1b}. Therefore, the likelihood used now has the full form of Eq. \ref{eq:likelihood}. 

In Figs. \ref{fig:meanc} and \ref{fig:stdc} we show the mean and standard deviation of the angular distance error of the inferred UHECR arrival direction using the samples that were produced through the updated posterior, conditional to both local POS data, as well as integrated LOS data. We observe that in comparison to scenario A, shown in Figs. \ref{fig:meana} and \ref{fig:stda}, the improvement in the ability to reconstruct the UHECR arrival direction is substantial in that the maximum mean angular error is reduced by a factor of about $1.5$, the part of the POS where the maximum mean angular error occurs is greatly reduced, and the variance of the posterior is diminished by a factor of about $1.2$. Thus, for the setting considered, we have shown that inclusion of integrated LOS data of the GMF - which is a much more realistic expectation compared to full 3D local measurements of scenario B - does also lead to significantly better results with regards to recovering the arrival directions of UHECRs with rigidity $r_*$.

\section{Discussion} \label{sec:discussion}

\subsection{Identification of a systematic bias} \label{sec:bias}

In Fig. \ref{fig:1a} we observe that the ordered component of the field primarily lies (anti)parallel to the $\pm \hat{y}$ direction, which corresponds to $l = \pm 90^\circ$ longitude. In Fig. \ref{fig:noRecon} this is reflected by the fact that the observed arrival directions parallel to the ordered component, $(l, b) \simeq (\pm 90^\circ, 0^\circ)$, are minimally deflected, while the maximal deflection occurs at the arrival directions perpendicular to the ordered component of the field. We call the map of Fig. \ref{fig:noRecon} the `deflection map' of the GMF, for a UHECR with rigidity $r_*$. If the deflection map of the GMF for a given of rigidity was available, then we would be able to identify the regions of the celestial sphere where observed UHECRS with that rigidity are deflected most.  

A comparison with Fig. \ref{fig:noRecon} with Fig. \ref{fig:mean} yields a direct correlation between the regions of the deflection map, and the mean angular error of our inferred arrival directions as a function of observed arrival direction, for the same rigidity. In qualitative terms, this correlation suggests that for observed arrival directions perpendicular to the GMF zero mode, where the particles must have deflected the most, our inference of their true arrival direction is more prone to a systematic bias. This `bias' is to be understood as the angular distance of the mean of our posterior distribution with respect to the true value.

Even though we might not be able to correct for this bias using our available data, knowledge of how severely the GMF alters the UHECR trajectories can help characterise the regions of the POS where our reconstructions are expected to suffer from it. While the corresponding deflection of the true GMF for a value of the UHECR rigidity will not be known a priori\footnote{Its derivation requires knowledge of the full $3$D structure of the GMF, which is unknown.}, its structure is largely dictated by the field's dominating mean value which is generally well captured by our algorithm as shown in \paperI. Indeed, as shown in Figs. \ref{fig:deflectionMapFFF} through \ref{fig:deflectionMapFFT}, we are able to recover the large-scale features of the deflection map accurately for all three considered scenarios, thus providing a charting of the parts of the POS where the GMF will most influence the UHECR trajectories, and by extension the regions where our arrival direction posterior might be shifted with respect to the true value.

\subsection{Caveats}

While tomography using starlight polarisation and Gaia data can provide the location of dust clouds in the local Galaxy as well as the POS orientation of the GMF at each cloud's location, the POS direction of the GMF is generally not known, as this inference makes use of the properties of grain alignment which cannot infer the POS directionality of the GMF (\citealt{Pasiphae}). %In real applications, this information can be supplemented from already existing GMF models such as the Jansson-Farrar 2021 (\citealt{JF12};\citealt{JF12-2}).

Further, the integrated measurements used here assume that the integrated Galactic LOS component has been measured or inferred. In practice, the observables that need to be measured in order to estimate these integrals is the Faraday rotation measure and the dispersion measure. That means that even if the Galactic component is separated, it will still provide an average weighted over the thermal electron density. Therefore, in our study we practically made the simplifying assumption that the thermal electron density is constant or known. In applications to the real GMF, the electron density will be treated as an additional degree of freedom to be inferred (\citealt{faraday}). However, it must be noted that recent research suggests the possibility that local LOS data can be available, at least in part of the dataset (\citealt{Tahani2022a}; \citealt{Tahani2022b}).

In this analysis we studied only the case of UHECRs with a fixed rigidity of $r_* = 5 \times 10^{19}$ eV. This is equivalent to assuming that the UHECRs particles are protons of $E = 5 \times 10^{19}$ eV. In general the composition of UHECRs is unknown, and is most likely mixed - especially if some of the sources have Galactic origin (\citealt{Kusenko1}; \citealt{Kusenko2}; \citealt{UHECRcomposition}). The closer examination of different composition scenarios will be the subject of future work.

\subsection{Conclusions \& Outlook}

In this paper we extended the analysis of \paperI $\text{}$ to the case of more realistic LOS information and local data distribution. This is motivated by the fact that in real applications, the local GMF data obtained through stellar polarisation tomography will not contain LOS information, and the distribution of these measurements will follow the distribution of dust clouds which is not homogeneous, as was assumed in \paperI.

Additionally, the ground-truth GMF that was used in order to benchmark the performance of our inference algorithm was taken from an MHD simulation, with the aim of studying the effect of our Gaussian approach to magnetic field configurations whose statistical properties more closely resemble those of the real GMF. Furthermore, we supplemented the existing framework in order to include LOS- integrated information as well. 

Our results show that while the complete absence of LOS information in the local data diminishes the accuracy of our inferred UHECR arrival directions, even in this case we are able to significantly correct for the effect of the GMF on the observed arrival directions, at least for the rigidity considered here. Yet, the inclusion of integrated LOS data for the GMF - which can be realistically expected to be part of our available information - is enough to provide accurate enough results.

Even in directions where the angular distance between the inferred arrival direction and the true are maximal, we are still able to correct for the effect of the GMF by a factor of $3$, in the setting considered. Additionally, by our ability to  reconstruct the large scale features of the field which dominate UHECR deflection, we are able to identify the regions of the POS where our reconstructions are most likely to have summer from maximal error.

\begin{acknowledgements}
 A.T. and V.P. acknowledge support from the Foundation of Research and Technology - Hellas Synergy Grants Program through project MagMASim, jointly implemented by the Institute of Astrophysics and the Institute of Applied and Computational Mathematics. A.T. acknowledges support by the Hellenic Foundation for Research and Innovation (H.F.R.I.) under the ``Third Call for H.F.R.I. Scholarships for PhD Candidates'' (Project 5332). V.P. acknowledges support by the Hellenic Foundation for Research and Innovation (H.F.R.I.) under the ``First Call for H.F.R.I. Research Projects to support Faculty members and Researchers and the procurement of high-cost research equipment grant'' (Project 1552 CIRCE). The research leading to these
results has received funding from the European Union’s Horizon 2020 research and innovation programme under the Marie Skłodowska-Curie RISE action, Grant
Agreement n. 873089 (ASTROSTAT-II). This work also benefited greatly from discussions during the program "Towards a Comprehensive Model of the Galactic Magnetic Field" at Nordita in April 2023, which is partly supported by NordForsk and the Royal Astronomical Society. A.T. would like to thank Vincent Pelgrims, Raphael Skalidis, Georgia V. Panopoulou, and Konstantinos Tassis for helpful tips and stimulating discussions. G.E. acknowledges the support of the German Academic Scholarship Foundation in the form of a PhD scholarship ("Promotionsstipendium der Studienstiftung des Deutschen Volkes"). P.F. acknowledges funding through the German Federal Ministry of Education and Research for the project ErUM-IFT: Informationsfeldtheorie fuer Experimente an Großforschungsanlagen (Foerderkennzeichen: 05D23EO1)
\end{acknowledgements}

\begin{appendix}\label{appendix:groundTruth}
\section{Simulated Magnetic Field} \label{sec:groundTruth}

We briefly summarize the setup and results of the Galactic dynamo simulations that have been analyzed here. A detailed description of the numerical setup is presented in \cite{bendre_2015}.

\begin{figure*}
\centering
    \includegraphics[width=0.9\linewidth]{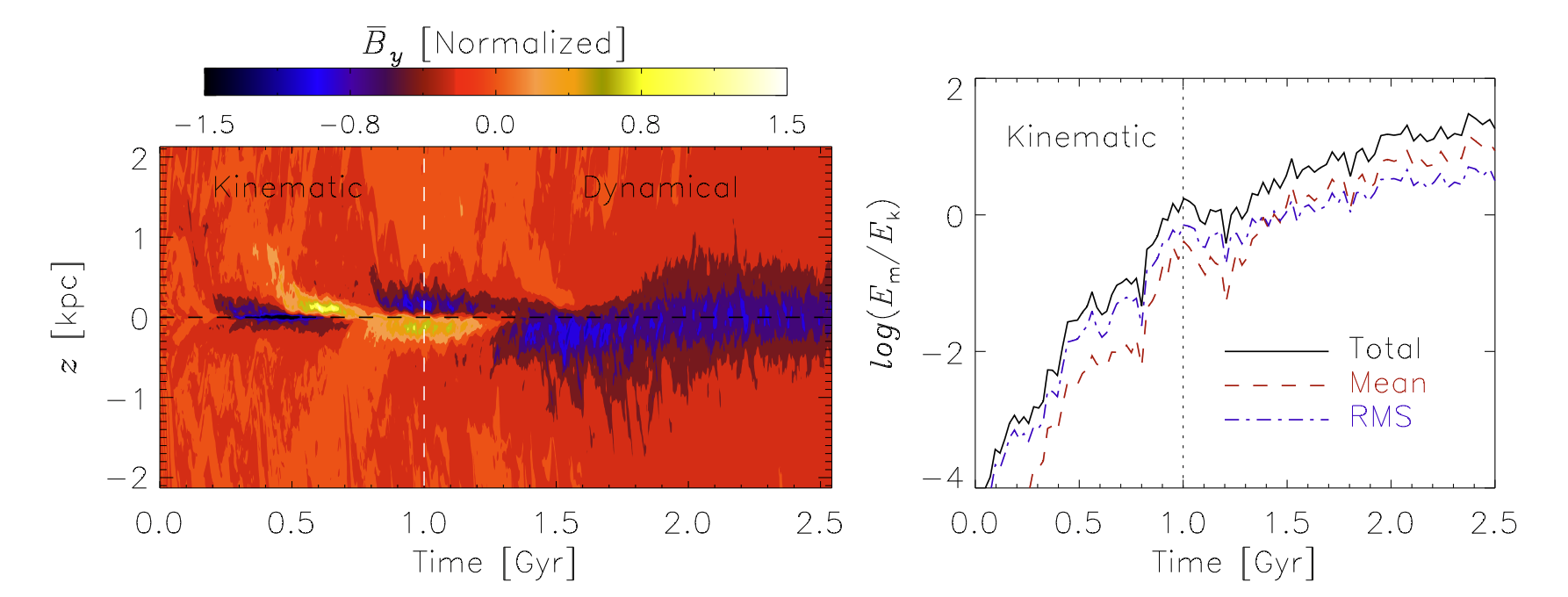}
    \caption{\textbf{Left}: Time evolution of the 
    vertical ($z$) profile of the azimuthal component of the 
    magnetic field averaged over $x-y$ plane. The color code is
    normalized by an exponential factor to compensate for an 
    exponential growth of magnetic fields. The mean 
    magnetic field eventually grows to a large-sale mode
    symmetric with respect to the Galactic mid-plane. \textbf{Right}: Time evolution of various 
    contributions to magnetic energy, normalized to the 
    turbulent kinetic energy (which stays roughly constant in 
    time). The black solid line corresponds to the total 
    magnetic energy contribution, the red dashed line corresponds 
    to the magnetic energy of mean magnetic fields (averaged 
    over the horizontal $x-y$ planes) and with the blue 
    dot-dashed line to the magnetic energy in the RMS magnetic 
    fields. The magnetic energy is amplified 
    exponentially for about a Gyr and eventually reaches an 
    equipartition with turbulent kinetic energy.}
    \label{fig:simulations_results}
\end{figure*}

These are Magnetohydrodynamic (MHD) simulations of the Galactic interstellar medium (ISM). The simulation domain is an elongated box, located roughly at the solar neighbourhood of the Milky Way. It has dimensions of approximately $1\times1$ kpc in the radial ($x$) and azimuth ($y$) direction and ranges from approximately $-2\, {\rm to\,}+2$ kpc in $z$ direction, on either side of the Galactic mid-plane. It is split in a uniform Cartesian 
grid with a resolution of approximately $8.3$ pc, and a set of non-ideal MHD equations is solved in this domain using the \texttt{NIRVANA} code \citep{ziegler} (see Eq. 1 from \cite{bendre_2015} for the set of equations we have solved). Periodic boundary conditions were used in the $y$ direction to incorporate the axisymmetry of the Galactic disc. The flat rotation curve is incorporated by allowing the angular velocity to scale inversely with the Galactic radius as $\Omega \propto1/R$, with $\Omega_0=100$ km s$^{-1}$ kpc$^{-1}$ at the centre of the box. Shearing periodic boundary conditions are used in the radial $x$ direction to accommodate the aforementioned radial dependence of angular velocity. The initial density distribution of the ISM is in hydrostatic balance with the vertical gravity pointing towards the mid-plane, such that the vertical scale-height of the initial density was approximately $300$ pc, with its value in the mid-plane of approximately $10^{-24}$ g cm$^{-3}$. A vertical profile of gravitational acceleration is adapted from \cite{gilmore}. The ISM in this box is stirred by supernovae (SN) explosions, which inject the thermal energy at random locations, at a rate of approximately $7.5$ kpc$^{-2}$ Myr$^{-1}$. The vertical distribution of the explosions scale with the mass density. A piece-wise power law, similar to \cite{sanchez-salcedo}, is used to model the temperature-dependent rate of radiative heat transfer, which along with SN explosions, roughly capture the observed multi-phase morphology of the ISM. We started the simulations with negligible initial magnetic fields of strength of the order of nG, and it grew exponentially to the strengths of the order of $\mu$G, with an e-folding time of about $200$ Myr, such that the final energy density of the magnetic fields reached to the equipartition with the kinetic energy density of the ISM turbulence (shown in the right-hand panel of Fig. \ref{fig:simulations_results}). The exponential amplification of the magnetic energy saturated after about a Gyr, and coherent magnetic fields of scale-height close to $500$pc were sustained in the box, consistent with the typical scale-height of GMFs (shown in the left-hand panel of Fig. \ref{fig:simulations_results}). The growth and saturation of these large-scale fields are understood in terms of a self-consistent large-scale dynamo mechanism, governed by the SN-driven stratified helical turbulence and the Galactic differential rotation \citep{bendre_2015}. 

\end{appendix}

% WARNING
%-------------------------------------------------------------------
% Please note that we have included the references to the file aa.dem in
% order to compile it, but we ask you to:
%
% - use BibTeX with the regular commands:
%   \bibliographystyle{aa} % style aa.bst
%   \bibliography{Yourfile} % your references Yourfile.bib
%
% - join the .bib files when you upload your source files
%-------------------------------------------------------------------
\bibliographystyle{aa}
\bibliography{bibliography.bib}
\end{document}